\documentclass[sigconf]{acmart}

\usepackage{threeparttable}
\usepackage{subcaption}
\usepackage{todonotes}
\usepackage{color, colortbl}

\usepackage{graphics}  
\usepackage{subcaption}
\usepackage{soul}
\usepackage{multirow}
\usepackage{adjustbox}
\usepackage{graphicx}
\usepackage[figuresright]{rotating}
\usepackage[labeled]{multibib}

\newcites{L}{Ludography}

\newcommand{\changed}{}

\AtBeginDocument{%
  \providecommand\BibTeX{{%
    \normalfont B\kern-0.5em{\scshape i\kern-0.25em b}\kern-0.8em\TeX}}}

\copyrightyear{2021}
\acmYear{2021}
\setcopyright{acmcopyright}\acmConference[CHI '21]{CHI Conference on Human Factors in Computing Systems}{May 8--13, 2021}{Yokohama, Japan}
\acmBooktitle{CHI Conference on Human Factors in Computing Systems (CHI '21), May 8--13, 2021, Yokohama, Japan}
\acmPrice{15.00}
\acmDOI{10.1145/3411764.3445307}
\acmISBN{978-1-4503-8096-6/21/05}

\begin{document}

\title{Player-AI Interaction: What Neural Network Games Reveal About AI as Play}


\author{Jichen Zhu}
\authornote{These authors contributed equally to this research.} \authornote{Currently at IT University of Copenhagen, Denmark.}
\email{jichen.zhu@gmail.com}
\affiliation{%
 \institution{Drexel University}
 \streetaddress{Philadelphia, PA, USA}
}
\author{Jennifer Villareale}
\authornotemark[1]
\email{jmv85@drexel.edu}
\affiliation{%
 \institution{Drexel University}
 \streetaddress{Philadelphia, PA, USA}
}
\author{Nithesh Javvaji}
\authornotemark[1]
\email{javvaji.n@northeastern.edu}
\affiliation{%
 \institution{Northeastern University}
 \streetaddress{Boston, MA, USA}
}
\author{Sebastian Risi}
\email{sebr@itu.dk}
\affiliation{%
 \institution{IT University Copenhagen}
 \streetaddress{Copenhagen, Denmark}
}
\author{Mathias Löwe}
\email{malw@itu.dk}
\affiliation{%
 \institution{IT University Copenhagen}
 \streetaddress{Copenhagen, Denmark}
}
\author{Rush Weigelt}
\email{rw643@drexel.edu}
\affiliation{%
 \institution{Drexel University}
 \streetaddress{Philadelphia, PA, USA}
}
\author{Casper Harteveld}
\authornotemark[1]
\email{c.harteveld@northeastern.edu}
\affiliation{%
 \institution{Northeastern University}
 \streetaddress{Boston, MA, USA}
}

\renewcommand{\shortauthors}{Zhu, Villareale and Javvaji, et al.}

\begin{abstract}
The advent of artificial intelligence (AI) and machine learning (ML) bring human-AI interaction to the forefront of HCI research. This paper argues that games are an ideal domain for studying and experimenting with how humans interact with AI. Through a systematic survey of neural network games (\textit{n} = 38), we identified the dominant interaction metaphors and AI interaction patterns in these games. In addition, we applied existing human-AI interaction guidelines to further shed light on player-AI interaction in the context of AI-infused systems. Our core finding is that {\em AI as play} can expand current notions of human-AI interaction, which are predominantly productivity-based. In particular, our work suggests that game and UX designers should consider flow to structure the learning curve of human-AI interaction, incorporate discovery-based learning to play around with the AI and observe the consequences, and offer users an invitation to play to explore new forms of human-AI interaction. 

\end{abstract}

\ccsdesc[500]{Human-centered computing~Human computer interaction (HCI)}
\ccsdesc[500]{Applied computing~Computer games}
\ccsdesc[300]{Computing methodologies~Artificial intelligence}

\keywords{Human-AI Interaction; Neural Networks; User Experience; Game Design}


\begin{teaserfigure}
 \includegraphics[width=\textwidth]{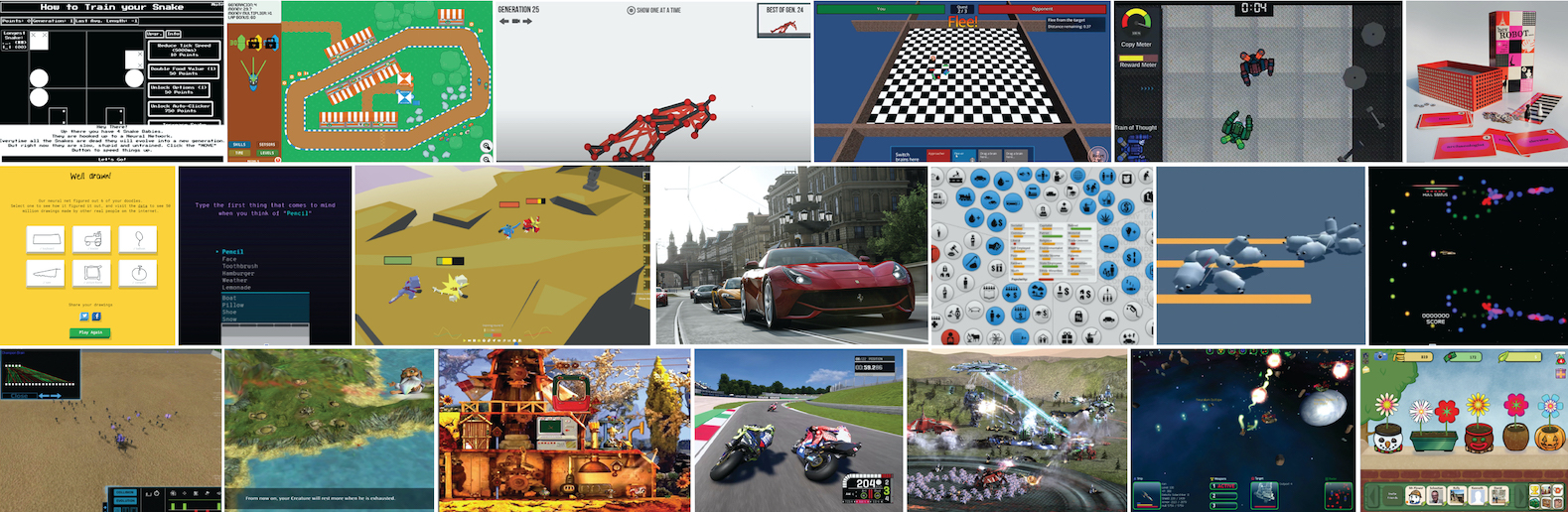}
 \caption{Selection of neural network (NN) games.  (From left to right, Top: \textit{How to Train Your Snake}~\protect\citeL{snake}, \textit{Idle Machine Learning Game}~\protect\citeL{idle}, \textit{Evolution}~\protect\citeL{evolution}, \textit{EvoCommander}~\protect\citeL{jallov2016evocommander}, \textit{Machine Learning Arena}~\protect\citeL{ferguson2019machine}, \textit{Hey Robot}~\protect\citeL{hey}; Middle: Quick, Draw!~\protect\citeL{quick}, \textit{Semantris}~\protect\citeL{semantris}, \textit{Dr. Derks Mutant Battlegrounds}~\protect\citeL{derks}, \textit{Forza Car Racing}~\protect\citeL{forza}, \textit{Democracy 3}~\protect\citeL{democracy}, \textit{Darwin's Avatars}~\protect\citeL{lessin2015darwin}, \textit{AudioinSpace}~\protect\citeL{hoover2015audioinspace}; Bottom: \textit{NERO}~\protect\citeL{stanley2005evolving}, \textit{Black \& White}~\protect\citeL{blackandwhite}, \textit{Creatures}~\protect\citeL{grand1997creatures}, \textit{MotoGP19}~\protect\cite{moto}, \textit{Supreme Commander 2}~\protect\citeL{rabin2015game}, \textit{Galactic Arms Race}~\protect\citeL{hastings2009evolving}, \textit{Petalz}~\protect\citeL{risi2015petalz})} 
\label{fig:games}
\end{teaserfigure}

\settopmatter{printfolios=true}

\maketitle

\section{Introduction}
With the recent boom in artificial intelligence (AI) technology, \footnote{Unless otherwise specified, we use the term AI broadly to include a wide range of artificial intelligence and machine learning techniques.} people are interacting with a growing number of AI-infused products in many aspects of everyday life. Already, these AI systems influence our decisions (e.g., recommendation systems~\cite{smith2017two}), inhabit our households (e.g., robotic appliances~\cite{sung2007my}), and accompany us in our playful experiences~\cite{mateas2003faccade,zhu2010towards,zhu2013shall} and educational games~\cite{valls2015exploring,zhu2019programming}. 
%

The technological development has precipitated renewed interest in the Human Computer Interaction (HCI) community. In addition to improving the usability of individual products~\cite{myers2018patterns}, HCI researchers synthesized design guidelines for human-AI interaction from the past decades~\cite{amershi2019guidelines}. There is a growing recognition that, compared to traditional interactive systems, AI-infused products impose additional challenges (e.g., technical barrier, low interpretability) to the current user experience (UX) design process~\cite{dove2017ux}. 
Furthermore, new interdisciplinary research areas have emerged around topics such as explainable AI~\cite{tintarev2007survey,zhu2018explainable,binns2018s,rader2018explanations}, ethics \& fairness~\cite{bryson2010robots, dove2017ux,holmquist2017intelligence}, and machine learning (ML) as a design material for UX~\cite{yang2018investigating,yang2018mapping}.

Among the fast-growing body of literature on human-AI interaction in the CHI community, one overlooked area is the context of play. With few exceptions, most recent literature focuses on productivity-related domains such as e-commerce, navigation and autocomplete~\cite{amershi2019guidelines}. While these are important domains for human-AI interaction, the history of AI and human-AI interaction has long been associated with play. For instance, {\em ELIZA}, one of the first AI programs designed to interact with lay users, was a playful satire of a certain school of psychotherapy~\cite{weizenbaum1966eliza}. Games such as {\em Chess}, {\em Poker}, {\em Go}, and {\em StarCraft} have continued to serve as benchmarks that propelled the development of AI since the beginning of the field. Since games naturally focus on end-user experience, game AI research has accumulated valuable knowledge related to human-AI interaction~\cite{mateas1999oz,mateas2003faccade,young2004architecture,zhu2008daydreaming,risi2015neuroevolution,valls2017graph}. 

In this paper, we propose the new construct of {\em player-AI interaction} to highlight how people interact with AI in the context of play, especially through computer games.  
To provide an overview of existing work in this area, we conducted the {\em first systematic review of player-AI interaction} in the scope of {\em Neural Network games} --- computer games in which players interact with an NN as part of the core gameplay. \changed{A neural network (NN) is a computational model that includes nodes (i.e., neurons) and connections between these nodes that transmit information \cite{stanley2007compositional}. The strengths of these connections (i.e., weights) are typically adjusted through some learning process. For our paper, this definition covers both NNs that control agents / non-player characters (NPCs) in games~\cite{evocommander:jallov2016evocommander,vinyals2019grandmaster} and generative NN models that produce game content~\cite{gar:hastings2009evolving,risi2015neuroevolution}.

\changed{We chose NN games for two key reasons. First, given the wide adoption of AI in games, we had to constrain our systematic (qualitative) review. 
Second, and more importantly, NN games provide insights into some of the most pressing open problems in human-AI interaction. For example, NNs are notorious for UX designers to work with because of NNs' low interpretability of the underlying process and the frequent unpredictability of its outcome}. Studying NN games can thus provide valuable information on how game designers have to work with these challenges.} 
 

We collected 38 NN games and applied a two-phased qualitative analysis to examine them. In the first phase, we use close reading and grounded theory to identify the overarching interaction metaphors and patterns of how NNs are represented in the game user interface (UI). In the second phase, we apply current human-AI interaction design guidelines~\cite{amershi2019guidelines}, compiled from a wide range of productivity-based domains, to our dataset. From these analyses, we derive design lessons for where games do well and identify open areas that can expand our current notion of human-AI interaction. A key design insight is that reframing {\em AI as play} offers a useful approach for considering human-AI interaction in games and beyond. 


The core argument of this paper is that games are a rich and currently overlooked domain for advancing human-AI interaction. The design space afforded by structuring AI as play, as game designers have been exploring, can point out new opportunities for AI-infused products in general. At the same time, insights of the generalized guidelines from other domains can be adapted to improve player-AI interaction. 
The key contributions of this paper are as follows: 
\begin{itemize}
    \item We propose the new research area of player-AI interaction. Through the \changed{first systematic review} on player-AI interaction in the context of NN games, we showcase how player-AI interaction can expand the current productivity-based discussions around human-AI interaction. 
    \item We adapted existing design guidelines for human-AI interaction to the context of games. Currently, there are no synthesized metrics to evaluate player-AI interaction. 
    \item We provide several insights from NN games (e.g., flow, exploration) to improve current challenges in human-AI interaction (e.g., learnability of AI).
\end{itemize}

\section{Related Work}
In this section, we summarize related work in human-AI interaction and AI-based games research. We aim to bridge these two disconnected areas. 

\subsection{Human-AI Interaction}
\label{sect:RelatedWork_HAI}

The HCI community has developed a body of work on how to design user interactions to improve productivity through AI-based applications~\cite{herlocker2000explaining,horvitz1999principles,hook2000steps,steinfeld2006common,winograd2006shifting}. 
Thanks to increasingly sophisticated big data and deep neural networks (i.e., deep learning), AI-infused products have started to enter the consumer market, prompting a new surge of interest in human-AI interaction~\cite{amershi2019guidelines,bansal2019beyond,antti2020GUI} and its societal impact in topics such as explainability and transparency~\cite{tintarev2007survey,zhu2018explainable,binns2018s, rader2018explanations}. 

An important research area is user-centered design for human-AI interaction. For instance, in conversational agents, a widely adopted type of AI-infused products, researchers have reported the wide gap between user expectations and the user experience (UX) of these systems~\cite{luger2016like} and how users develop their own strategies to work around the obstacles~\cite{myers2018patterns}. 
\changed{Recently, researchers synthesized guidelines, principles, and theories into coherent design frameworks for human-AI interaction~\cite{amershi2019guidelines,sukis2019ai,wang2019designing}.}
While user-centered (or player-centered) design is key in game development~\cite{sweetser2004player}, few works have looked at games as a domain for human-AI interaction, despite the long history of games and AI.  
Our work is the first to do so in a systematic and empirical way. More specifically, we leverage the existing meta-review of human-AI interaction design principles ~\cite{amershi2019guidelines} to investigate NN games.

There is a growing understanding in recent literature that designing AI and ML products is especially challenging for UX designers. \citeauthor{dove2017ux}~\cite{dove2017ux} acknowledged that, despite the regular use of ML in UX products, there has been little design innovation. Echoing the challenges of using ML as design material, \citeauthor{yang2020HAI}~\cite{yang2020HAI} argued that capability uncertainty and output complexity of AI systems are the two root causes of why human-AI interaction is uniquely tricky to design. The work presented in this paper aims to identify player-AI interaction, which is currently separated from the mainstream human-AI interaction literature, as a rich domain for further study and experimentation. \changed{Lessons from the game research community on how to structure human-AI interaction in the context of play can help to expand the current body of work in human-AI interaction.}

\subsection{AI-based Game Design and Player Experience}
\label{sect:RelatedWork_games}
In AI research, there is an extended history of using games as a rich domain to motivate algorithmic advancements. Salient examples include {\em Chess} in the era of ``Good Old Fashioned AI'' (GOFAI)~\cite{campbell2002deep}, {\em Go}~\cite{silver2016mastering}, classic Atari video games~\cite{mnih2013playing}, or even the popular AAA game {\em StarCraft}~\cite{ontanon2013survey, vinyals2019grandmaster} in the age of deep learning. The advances in game AI in turn opened new design spaces of player experience in research~\cite{mateas1999oz,mateas2003faccade,young2004architecture,zhu2008daydreaming,valls2017graph} as well as commercially released games and game engines (e.g., {\em Versu}~\cite{evans2013versu}, {\em Left4Dead}~\cite{valve2008left4dead} and {\em Civilization VI}~\cite{civil6}).


However, with few exceptions, games have only recently started to be used as a serious domain for human-AI interaction research. For instance, \citeauthor{Gomme2020}~\cite{Gomme2020} used strategy games to study players' expectations for what they consider to be a worthy AI-controlled opponent. Along those lines, several researchers proposed using games and playful experiences to help designers and users learn AI~\cite{Myers20_QUBE,fulton2020getting,pemberton2019}.   

Most existing work has focused on high-level metaphors (often referred to as ``design patterns'') of how players and designers can interact with AI. For instance, Treanor et al.~\cite{treanor2015ai} derived nine patterns based on what players do: AI as role-model, trainee, editable, co-creator, adversary, villain, or spectacle, and whether AI is visible or guided. Cook et al.~\cite{cook2016pcg} further examined design patterns in procedural content generation (PCG)-based games and derived different AI design patterns. 
In the context of assisting the game development process, Riedl and Zook~\shortcite{riedl2013ai} proposed that AI plays the role of actor, designer, and producer. While the above work provides a critical starting point for our work, they are ``meant to be a tool for thinking about creating AI-based games, rather than serve as a comprehensive taxonomy of methods''~\cite{treanor2015ai}. Finally, \citeauthor{guzdial2019friend}~\cite{guzdial2019friend} used the taxonomy of friend, collaborator, student, and manager to describe the different interaction metaphors for how game designers interact with an AI-based game level editor. Our work builds on this tradition of using human-human interaction as metaphors to structure the interaction between humans and AI. In addition, we extend this literature by conducting an in-depth empirical analysis through grounded theory instead of relying on researchers' domain expertise, as is the case for the above-mentioned existing work. 


Finally, there is a significant body of work in games research to understand player experience \cite{nacke2009playability,lucero2013playful,desurvire2009game,denisova2015,abeele2020development}. For example, the game engagement questionnaire (GEQ)~\cite{brockmyer2009development} is a widely used instrument for measuring player engagement, although recently it has been approached with increasing criticism~\cite{law2018systematic}. Other notable frameworks include game involvement \cite{calleja2007digital}, game usability~\cite{desurvire2009game}, and design heuristics~\cite{lucero2013playful}. While these frameworks are useful to improve the general player experience, they do not have sufficient focus on the interaction between players and AI to guide the human-AI interaction design of games. Thus, our work proposes the first set of guidelines to design and evaluate player-AI interaction.

\section{Dataset: Neural Network Games}
\label{dataset}

This section describes our systematic search process and the resulting dataset of 38 NN Games. \changed{Table~\ref{tab:game_analysis_mainTable} provides an overview of this dataset, including the characteristics described in this section.} 

\subsection{Search Strategy and Data Collection}
\begin{figure}[!t]
 \includegraphics[width=0.5\textwidth]{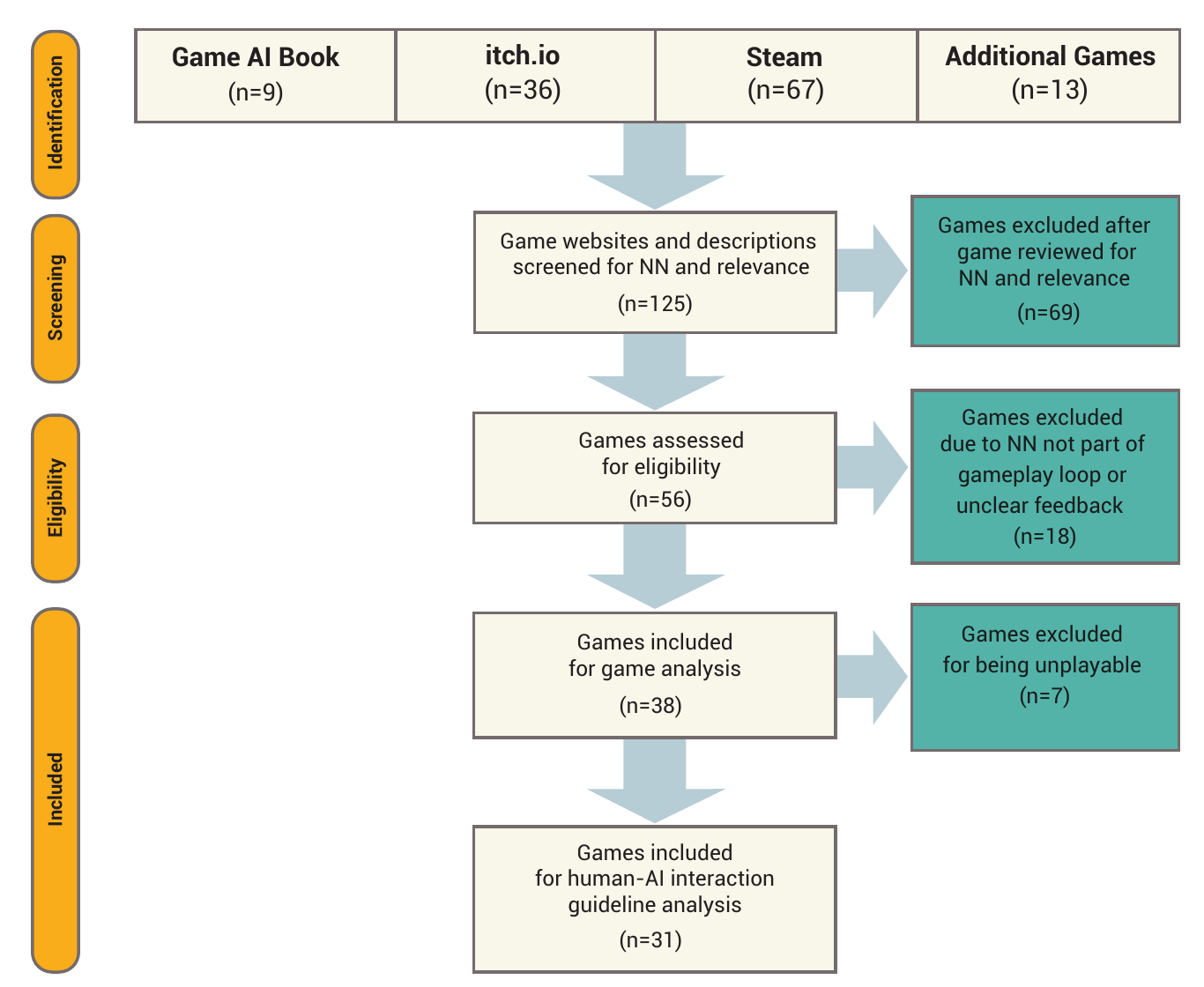}
 \caption{\changed{Data collection process.}} 
\label{fig:PRISMA}
\end{figure}

We searched two popular web gaming portals---{\em Steam} and {\em itch.io}---and a widely used game AI book, {\em Artificial Intelligence and Games}~\cite{yannakakis2018artificial}. We chose these three sources because they collectively cover a wide variety of games of different production modes. {\em Steam} is the largest digital distribution platform for PC gaming, offering over 34,000 games with over 95 million monthly active users in 2019~\cite{steamwiki}. {\em itch.io} is one of the largest platforms for indie games, containing nearly 100,000 games with various metadata. {\em Artificial Intelligence and Games} is the most cited book on game AI that includes examples of games with notable AI innovations. The book complements the previous two sources for its coverage on AAA commercial games and research games. 

Our inclusion criteria were computer games \changed{wherein players can interact with an NN as part of the core gameplay (i.e., gameplay loop). We use the definition of core gameplay as ``the set of actions the player iterates on the most while playing the game [and which] should directly influence the outcomes of the game''~\cite{guardiola2016gameplay}. We further excluded work with no clear win condition {\em and} no clear feedback on how player interaction with the AI impacts the game, as these games lack the basic elements for meaningful player-AI interaction. Notice that sandbox games with clear feedback to player interaction are included~\cite{blackandwhite,aidungeon,corral,creatures:grand1997creatures,gar:hastings2009evolving}. For the same reason, we excluded games where the NN did not interact with players (e.g., ML agents that can automatically play games~\cite{snodgrass2014experiments}). Finally, we excluded digitized versions of traditional board/card games (e.g., {\em Chess}, {\em Go}, {\em Poker}) to focus on computer games. Future research is needed for investigating player-AI interaction in traditional games.}


Our search process is summarized in Figure~\ref{fig:PRISMA}. Similar to other systematic reviews on large game repositories~\cite{alharthi2018playing}, we used pre-existing game tags in these systems. On {\em itch.io}, we used its tags ``neural-network'' and ``machine-learning,'' and ``AI.'' 
On {\em Steam} we searched with the terms ``neural network'', ``machine learning'', and also used {\em Steam}’s own tag ``artificial intelligence.'' We acknowledge that not all NN games identify themselves as such, and this is a limitation of our study. \changed{However, it is possible that the games that advertise their use of NNs are more likely to pay extra attention to player-AI interaction.} 
In the {\em Artificial Intelligence and Games} book~\cite{yannakakis2018artificial}, we went through \changed{the chapter ``A Brief History of Artificial Intelligence and Games'' to collect the relevant games.} 
In order to include as many influential examples as possible, we also asked on social media in the games and game AI communities for additional work. The suggested games are included in the category of ``additional games'' along with the games the authors were aware of. 

\changed{After screening the 125 games resulting from the above process for eligibility, we found 38 games that met the inclusion criteria (Table~\ref{tab:game_analysis_mainTable}).} The most common reasons for games to be excluded were that 1) they used content (e.g., music) generated by an external NN, but the NN was not part of the gameplay loop, and 2) they did not have full human-AI interaction due to the lack of feedback for player actions. For example, {\em Bird by Example} \citeL{bird} is a single-player RPG where players navigate a forest environment and interact with their bird offspring that is controlled by an NN. This game was excluded because the player's actions (i.e., walk, jump, punch) produced no visible change in the NN's behavior. As a result, it was unclear how the player was intended to interact with the NN and to what end the NN impacted gameplay. Note that our goal was not to develop a comprehensive list, but rather to capture a representative sample of NN games to analyze current trends in player-AI interaction. 



\subsection{Dataset}
For each game we included, two researchers collected the following to form our dataset: 1) screen recording of one researcher playing at least one hour of the game, 2) game developers' description of the game and their design intent (i.e., via the game's website, developer blog, and academic publications), and 3) technical features of the NN (online vs. offline learning, the types of output to the NN). We used findings from 1) and 2) to determine 3). If it was not apparent, the two researchers consulted our NN expert coauthors for resolution. This section summarizes the key characteristics of our dataset. \changed{It should be noted that seven games did not have playable versions publicly available (marked with * in Table~\ref{tab:game_analysis_mainTable}).} For those games, the researchers used existing gameplay footage available on the Internet for our Phase 1 analysis. However, for Phase 2, directly playing the games is necessary. 
Thus, we excluded the unplayable games for our guideline analysis (see Figure~\ref{fig:PRISMA}).

\subsubsection{Characteristics as Games}
As a collection of games, our dataset is diverse in multiple aspects. \changed{Indie and research games make up most of the dataset (74\%), while only 26\% are AAA games. Using ~\citeauthor{heintz2015game}'s taxonomy~\cite{heintz2015game}, our games cover all the genres: 15 simulation games (39\%), 7 puzzle games (18\%), 5 strategy games (13\%), 4 role-playing games (11\%), 3 action games (8\%), 3 sports games (8\%) and 1 adventure game (3\%). In addition, 16 games are multiplayer (42\%), and the remaining 22 games are single player (58\%).} \changed{This suggests that our dataset has a good representation of different types of games.}

\subsubsection{Characteristics of NN and Game AI}
From the technical point of view, our dataset also covered a wide range of varieties. \changed{There is a relatively even split between NN games with {\em online learning} (58\%) and {\em offline learning} methods (42\%).} This technical feature is associated with different gameplay characteristics. In online learning games, the network is (further) trained as the player interacts with it. Therefore, these games can adapt to individual players' actions in real-time. Offline learning games, on the other hand, are shipped with fixed NNs and are not adaptive in the same way. However, offline learning games have the advantage of handling more complex user input, such as natural language~\citeL{aidungeon,semantris,guesstheword,hey} and images~\citeL{villareale2020innk,quick}. 
%
The output of the NN can be divided into {\em behavior} (89\%) and {\em content} (11\%). \changed{Behavior output consists of the actions and decisions by NN-controlled characters. Content output typically take the form of in-game assets such as flowers~\citeL{risi2015petalz} or weapons~\citeL{hastings2009evolving}}.


\changed{A key challenge to our analysis is the black-box nature of AI and NNs, especially from the players' point of view. Similar to other AI-infused products, games often contain complex interactions supported by different algorithms. It can be difficult to attribute gameplay features to specific algorithms without access to source code. The authors, including two game AI researchers, made our best effort to determine whether there were multiple AIs in a game (e.g., {\em Black and White}~\citeL{blackandwhite}) and what the NN was responsible for. We did so by using game developers' descriptions (e.g., conference talks and online articles) and our technical expertise to analyze the gameplay. Still, many such technical details remain unknown for commercial games. We acknowledge this as a limitation of our study. However, what makes NNs uniquely demanding in their human-AI interaction design is their reduced predictability and low interpretability. We argue that an NN, whether as the entire game AI or as a component, will introduce these characteristics to the games they are part of. As a result, NN games can be studied as a whole, regardless of their technical differences.}


\begin{sidewaystable*}
\caption{Overview of the 38 NN games and the results of Phase 1 analysis. (* Games without available playable versions and excluded for Phase 2.)} 
\label{tab:game_analysis_mainTable}
\begin{adjustbox}{width=\textwidth}
\begin{tabular}{lccclcccc}
\rowcolor[HTML]{F3F3F3} 
\cellcolor[HTML]{F3F3F3}                                 & \multicolumn{2}{c}{\cellcolor[HTML]{F3F3F3}\textbf{Game Characteristics}} & \multicolumn{4}{c}{\cellcolor[HTML]{F3F3F3}\textbf{NN Characteristics}}                                                                                                & \multicolumn{2}{c}{\cellcolor[HTML]{F3F3F3}\textbf{Player-NN Framework Characteristics}}   \\
\cmidrule(r){2-3}\cmidrule(r){4-7}\cmidrule(r){8-9}
\multicolumn{1}{l}{\textbf{Game Title {[}Ludography{]}}} & \textbf{Publisher}                & \textbf{Genre}                        & \textbf{Multiple AIs?}          & \multicolumn{1}{c}{\textbf{NN Responsibilities}}               & \textbf{NN Output}                & \textbf{Learning}               & \textbf{Interaction Metaphor}      & \textbf{UI}                           \\
\rowcolor[HTML]{F3F3F3} 
2D Walk Evolution {\citeL{2dwalkevolution}}                                        & Indie                             & Simulation                            & No                              & controls creature movement                                     & Behaviors                         & Offline                         & Teammate                           & NN-Specific                           \\
AI Dungeon {\citeL{aidungeon}}                                              & Indie                             & Adventure                             & No                              & creates natural language responses                             & Behaviors                         & Offline                         & Designer                           & NN-Limited                            \\
\rowcolor[HTML]{F3F3F3} 
AIvolution {\citeL{aivolution}}                                              & Indie                             & Simulation                            & Unknown                         & controls creature movement                                     & Behaviors                         & Online                          & Teammate                           & NN-Limited                            \\
AudioinSpace* {\citeL{hoover2015audioinspace}\cite{audioinspace:hoover2015audioinspace}}                                           & Research                          & Action                                & Yes                             & creates weapon visuals and audio                               & Content                           & Online                          & Designer                           & NN-Agnostic                           \\
\rowcolor[HTML]{F3F3F3} 
Black \& White {\citeL{blackandwhite}\cite{blackandwhite:wexler2002artificial}}                                          & AAA                               & Role-play                             & Yes                             & creates creature desires                                       & Behaviors                         & Online                          & Apprentice                         & NN-Agnostic UI                        \\
Blitzkrieg 3 {\citeL{blitzkrieg}}                                            & AAA                               & Strategy                              & Unknown                         & controls "Boris" battle behavior                               & Behaviors                         & Offline                         & Competitor                         & NN-Limited UI                         \\
\rowcolor[HTML]{F3F3F3} 
BrainCrafter* {\cite{braincrafter}}                                          & Research                          & Puzzle                                & No                              & controls robot movement                                        & Behaviors                         & Online                          & Apprentice                         & NN-Specific UI                        \\
Colin McRae Rally 2.0* {\citeL{colin}}                                  & AAA                               & Sports                                & Unknown                         & \cellcolor[HTML]{F3F3F3}controls the car's driving performance & Behaviors                         & Offline                         & Competitor                         & NN-Agnostic UI                        \\
\rowcolor[HTML]{F3F3F3} 
Competitive Snake {\citeL{competitivesnake}}                                       & Indie                             & Puzzle                                & No                              & controls enemy snake behavior                                  & Behaviors                         & Offline                         & Competitor                         & NN-Agnostic UI                        \\
Corral {\citeL{corral}}                                                  & Indie                             & Simulation                            & Unknown                         & controls chicken movement and preservation skills              & Behaviors                         & Online                          & Apprentice                         & NN-Agnostic UI                        \\
\rowcolor[HTML]{F3F3F3} 
Creatures {\cite{grand1997creatures}\cite{creatures:grand1997creatures}}                                               & AAA                               & Role-play                             & Yes                             & controls the creature's sensor-motor coordination              & Behaviors                         & Online                          & Apprentice                         & NN-Agnostic UI                        \\
Darwin's Avatar* {\citeL{lessin2015darwin} \cite{darwinsavatar:lessin2015darwin}}                                       & Research                          & Action                                & No                              & controls creature movement                                     & Content                           & Offline                         & Designer                           & NN-Limited UI                         \\
\rowcolor[HTML]{F3F3F3} 
Democracy 3 {\citeL{democracy}}                                             & AAA                               & Role-play                             & Unknown                         & creates motivations and desires of the public                  & Behaviors                         & Offline                         & Designer                           & NN-Agnostic UI                        \\
Dr. Derk's Mutant Battlegrounds {\citeL{derks}}                         & Indie                             & Simulation                            & Unknown                         & controls creature movement and behavior                        & Behaviors                         & Online                          & Apprentice                         & NN-Limited UI                         \\
\rowcolor[HTML]{F3F3F3} 
EvoCommander* {\citeL{jallov2016evocommander} \cite{evocommander:jallov2016evocommander}}                                          & Research                          & Simulation                            & No                              & controls tank movement and shooting behavior                   & Behaviors                         & Online                          & Apprentice                         & NN-Specific UI                        \\
Evolution {\citeL{evolution}}                                               & Indie                             & Simulation                            & Unknown                         & controls creature movement                                     & Behaviors                         & Online                          & Teammate                           & NN-Specific UI                        \\
\rowcolor[HTML]{F3F3F3} 
evolution for beginners {\citeL{evolutionbeg}}                                 & Indie                             & Simulation                            & Unknown                         & controls creature movement and sensory input                   & Behaviors                         & Online                          & Apprentice                         & NN-Limited UI                         \\
Football Evo {\citeL{football}}                                           & Indie                             & Simulation                            & No                              & controls player movement and behavior                          & Behaviors                         & Online                          & Apprentice                         & NN-Limited UI                         \\
\rowcolor[HTML]{F3F3F3} 
Forza Car Racing {\citeL{forza}\cite{forza:takahashi_2018}}                                        & AAA                               & Sports                                & Unknown                         & controls the car's driving performance                         & Behaviors                         & Online                          & Competitor                         & NN-Limited UI                         \\
GAR {\cite{hastings2009evolving}\cite{gar:hastings2009evolving}}                                                     & Research                          & Action                                & Yes                             & creates particle weapons                                       & Content                           & Online                          & Designer                           & NN-Limited UI                         \\
\rowcolor[HTML]{F3F3F3} 
Gridworld {\citeL{grid}}                                               & Indie                             & Simulation                            & Unknown                         & controls creature behavior                                     & Behaviors                         & Online                          & Designer                           & NN-Limited UI                         \\
Guess the Word {\citeL{guesstheword}  \cite{guesstheword:publication}}                                          & Research                          & Puzzle                                & Yes                             & creates natural language responses                             & Behaviors                         & Offline                         & Teammate                           & NN-Limited UI                         \\
\rowcolor[HTML]{F3F3F3} 
Hey Robot {\citeL{hey}}                                               & Indie                             & Puzzle                                & No                              & controls language processing                                   & Behaviors                         & Offline                         & Teammate                           & NN-Agnostic UI                        \\
How to Train your Snake {\citeL{snake}}                                 & Indie                             & Simulation                            & No                              & controls snake movement                                        & Behaviors                         & Online                          & Apprentice                         & NN-Specific UI                        \\
\rowcolor[HTML]{F3F3F3} 
Idle Machine Learning Game {\citeL{idle}}                              & Indie                             & Simulation                            & No                              & controls performance of the vehicle's movement                 & Behaviors                         & Online                          & Apprentice                         & NN-Specific UI                        \\
iNNk {\citeL{villareale2020innk}}                                                    & Research                          & Puzzle                                & No                              & identifies sketches drawn by the player                        & Behaviors                         & Offline                         & Competitor                         & NN-Specific UI                        \\
\rowcolor[HTML]{F3F3F3} 
Machine Learning Arena* {\citeL{ferguson2019machine}}                                & Research                          & Simulation                            & Unknown                         & controls robot behavior                                        & Behaviors                         & Online                          & Teammate                           & NN-Specific UI                        \\
MotoGP19 {\citeL{moto}}                                                & AAA                               & Sports                                & Unknown                         & controls the car's driving performance                         & Behaviors                         & Offline                         & Competitor                         & NN-Agnostic UI                        \\
\rowcolor[HTML]{F3F3F3} 
Neat Race {\citeL{neat}}                                               & Indie                             & Simulation                            & No                              & controls car movement                                          & Behaviors                         & Online                          & Apprentice                         & NN-Specific UI                        \\
NERO {\citeL{stanley2005evolving}\cite{nero:stanley2005evolving}}                                                    & Research                          & Simulation                            & No                              & controls robot movement and shooting behavior                  & Behaviors                         & Online                          & Apprentice                         & NN-Specific UI                        \\
\rowcolor[HTML]{F3F3F3} 
Oui Chef!! {\citeL{oui}\cite{cimolino2019oui}}                                              & Research                          & Role-play                             & No                              & controls chef behavior                                         & Behaviors                         & Online                          & Apprentice                         & NN-Agnostic UI                        \\
Petalz* {\citeL{risi2015petalz}\cite{petalzpublication:risi2015petalz}}                                                 & Research                          & Simulation                            & No                              & creates flowers                                                & Content                           & Offline                         & Designer                           & NN-Agnostic UI                        \\
\rowcolor[HTML]{F3F3F3} 
Quick, Draw! {\citeL{quick}}                                            & Research                          & Puzzle                                & No                              & identifies sketches drawn by the player                        & Behaviors                         & Offline                         & Teammate                           & NN-Limited UI                         \\
Race for the Galaxy {\citeL{raceforthegalaxy}}                                     & AAA                               & Strategy                              & Unknown                         & controls opponent behavior                                     & Behaviors                         & Offline                         & Competitor                         & NN-Limited UI                         \\
\cellcolor[HTML]{F3F3F3}Roll for the Galaxy {\citeL{rollforthegalaxy}}             & \cellcolor[HTML]{F3F3F3}AAA       & \cellcolor[HTML]{F3F3F3}Strategy      & \cellcolor[HTML]{F3F3F3}Unknown & controls opponent behavior                                     & \cellcolor[HTML]{F3F3F3}Behaviors & \cellcolor[HTML]{F3F3F3}Offline & \cellcolor[HTML]{F3F3F3}Competitor & \cellcolor[HTML]{F3F3F3}NN-Limited UI \\
Semantris {\citeL{semantris}}                                               & Research                          & Puzzle                                & No                              & controls the classification of words                           & Behaviors                         & Offline                         & Teammate                           & NN-Limited UI                         \\
\rowcolor[HTML]{F3F3F3} 
Supreme Commander 2 {\citeL{rabin2015game}\cite{supremecommander:rabin2015game}}                                     & AAA                               & Strategy                              & Yes                             & controls enemy unit flight and fight behavior                  & Behaviors                         & Offline                         & Competitor                         & NN-Agnostic UI                        \\
The Abbattoir Intergrade {\citeL{abbattoir}}                                & Indie                             & Strategy                              & Unknown                         & controls enemy unit offense behavior                           & Behaviors                         & Online                          & Competitor                         & NN-Agnostic UI

\end{tabular}

\end{adjustbox}
\end{sidewaystable*} 

\section{Phase 1: Analyzing Player-AI Interaction in NN Games}
\label{sec:phase1}
The first broad question we attempt to answer is how existing games use neural networks (NN), especially in terms of human-AI interaction. In particular, we focused on two subsidiary research questions:
\begin{itemize}
    \item RQ 1.a: How do NN games structure player-AI interaction? 
    \item RQ 1.b: How visible are NNs in the UI of the core gameplay?
\end{itemize}

\subsection{Methods}
The overall analysis procedure involved a close reading of the gameplay data in our dataset of 38 games. We used grounded theory to iteratively develop a framework for how players interact with the NN (i.e., the interaction metaphors) and how much the existence of NN is foregrounded in the core gameplay (i.e., levels of visibility). 

After initial observations of the games, two researchers conducted a close reading of the dataset based on the following questions: 1) What role does the NN play in the overall game system? 2) How does the player interact with the NN in the game? 3) Where does the interaction with the NN occur in the gameplay experience? and 4) How, if at all, are the NNs presented in the UI? 
%


Next, the two researchers conducted a preliminary open coding to label notable characteristics of the observations. 
\changed{During this step of the analysis, both researchers first went through each game individually and noted initial labels (e.g., player input via parameters, NN outputs behavior, NN is represented as a creature) into a shared document. Then, they discussed this document to iterate on the labels to form concepts (e.g., player directs NN toward a desired goal).} 
While constructing the concepts, they re-observed some of the games and reviewed related literature to refine the classifications. 

Once a common concept list was achieved and agreed upon, both researchers separately re-analyzed the shared document to develop preliminary categories that fit into a framework. During this step of the analysis, the researchers presented each other's framework to one another and then collectively iterated on the categories to finalize the framework. The result of this phase is a set of categories that make up the player-NN framework (i.e., the interaction metaphors discussed in Section~\ref{sec:metaphors} and levels of visibility discussed in Section~\ref{sec:visibility}). Using the framework, each researcher independently coded the same 7 games (20\%), one from each genre. 
After a complete agreement on the codes, they then coded the rest of the games independently. When the codes were complete, both researchers reviewed each other's work to ensure there were no discrepancies. If there was a discrepancy, the game would be discussed and reviewed again by both researchers. 

\changed{We opted for this consensual qualitative approach~\cite{hill2012consensual} instead of inter-rater reliability (IRR) because analyzing NN games and their compliance with guidelines (next section) is complex. This is due to the incredibly varying contexts and nuances that need to be considered in this emerging but under-studied area. Consensus coding is generally more suited for small samples and for considering multiple viewpoints, which fits our study better than IRR~\cite{mcdonald2019reliability}. Specifically, in order to get such multiple viewpoints, the two researchers discussed above were a game researcher and an AI engineer. A much richer and deeper understanding is thus gained.}

\subsection{Results}
This section presents the results of our Phase 1 analysis. \changed{All classifications of each game according to our player-NN framework are presented in Table~\ref{tab:game_analysis_mainTable}.}

\subsubsection{Interaction Metaphors}
\label{sec:metaphors}
The HCI literature shows that interface metaphors (e.g., ``Desktop'' and ``Search Engine'') are ``useful ways to provide familiar entities that enable people readily to understand the underlying conceptual model [of a system] and know what to do at the interface''~\cite[p.78]{Sharp2019}. Critical AI studies revealed the importance of metaphors to AI~\cite{agre1997computation,mateas2003expressive,zhu2009intentional}. Our analysis found four interaction metaphors that provide familiar structures for players to interact with the AI:  NN as \textit{Apprentice}, \textit{Competitor}, \textit{Designer}, and \textit{Teammate}. This finding is consistent with recent work in the game design literature. Based on their expert knowledge and intuition, game developers discuss how interaction metaphors (often referred to as ``design patterns'') have been used in game design~\cite{treanor2015ai,cook2016pcg} and in game production~\cite{riedl2013ai}. Our analysis extends the existing literature by conducting the first empirical work that uses deep qualitative analysis to analyze the interaction metaphors. 



\begin{figure}
    \centering
    \includegraphics[width=0.8\linewidth]{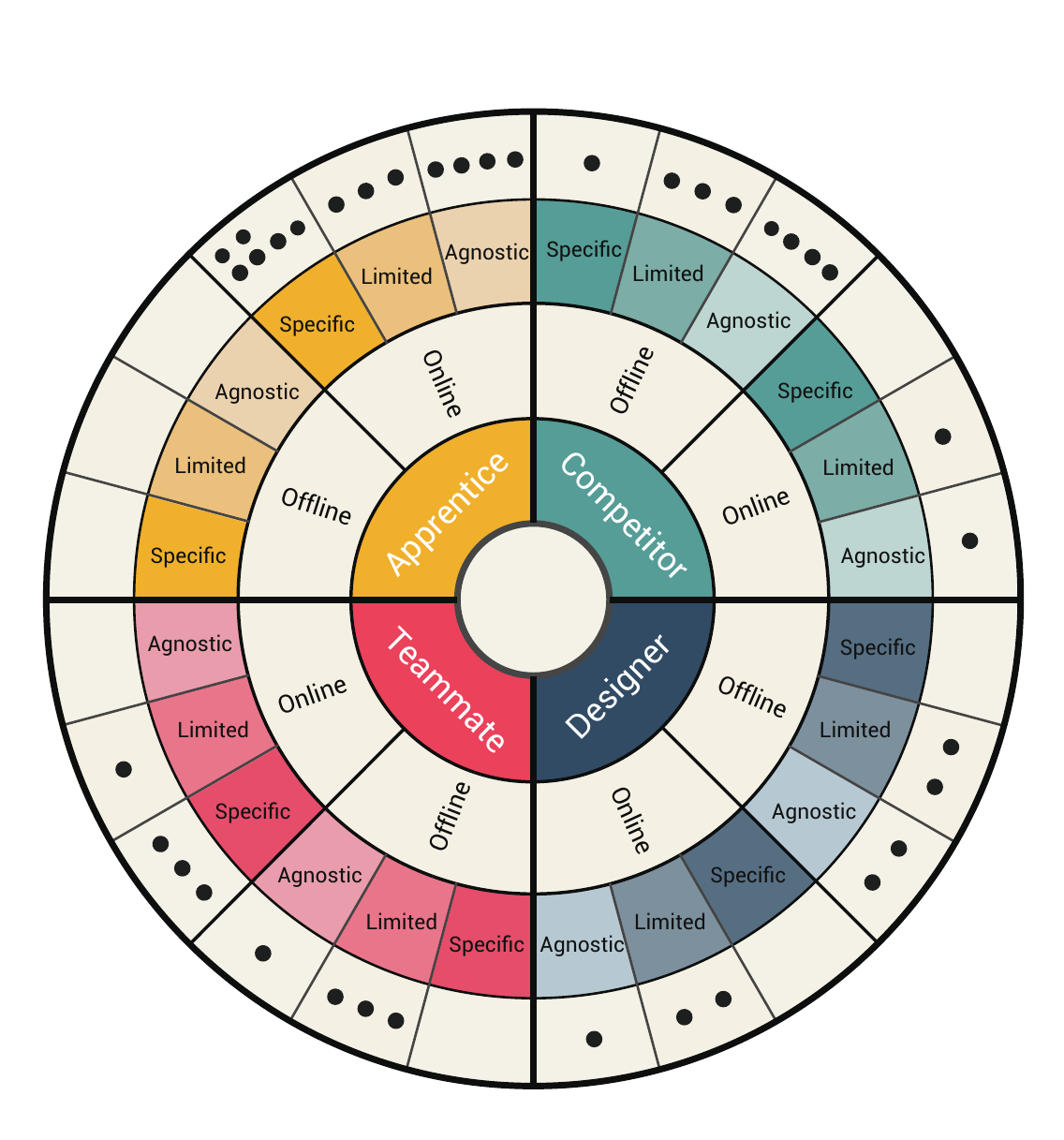}
    \caption{Distribution of the NN games (\textit{n} = 38) categorized by interaction metaphor, online/offline learning, and UI visibility. Each black dot represents one NN game.}
    \label{fig:game_analysis_circlechart}
\end{figure}

The largest portion of NN games \changed{(34\%) adopted what we call {\bf Neural Network as Apprentice}.} In these games, the player interacts with the NN as its mentor, and the focus of the gameplay is {\em how player changes the NN over time}. The player's mentoring of the NN can be achieved by providing direct feedback to the NN's behaviors~\citeL{blackandwhite,grand1997creatures,evolutionbeg}. For example, in {\em Creatures}, the player provides positive feedback (petting) when the NN-controlled creature displays desirable behavior (e.g., eat when hungry) and punishes it (slapping) for the opposite. A second way the player can mentor the NN is by configuring the right training setting for it~\citeL{football,hey,braincrafter,stanley2005evolving}. 
The gameplay afforded by this interaction metaphor focuses on getting the player to train the NN. As shown in Figure~\ref{fig:game_analysis_circlechart}, all games in this category use online learning.

Another interaction metaphor our NN games use is {\bf Neural Network as Competitor}. The key characteristics of this group, consisting of \changed{26\% of the games, is that player-AI interaction is adversarial.} For example, in \textit{Supreme Commander 2}~\citeL{rabin2015game}, the player fights an NN through their respective army platoons. As the player customizes their army, the NN weighs the player's unit composition against its own and makes tactical battle decisions, such as how its army will respond, which enemy to target first, or when to retreat. 
%
The NN can exploit players that are over-reliant on a single strategy and counter the player to create an evolving challenge~\citeL{forza,moto,rabin2015game,abbattoir,blitzkrieg,raceforthegalaxy,rollforthegalaxy}.
In these games, the NN counters the player during gameplay, thus encouraging them to adapt and try new strategies. A key distinction in this category is that the NN learns player's actions to create a more difficult challenge for the player to overcome. As discussed further in the next section, only one game~\citeL{villareale2020innk} here explicitly highlights the existence of the NN in their core UI. 


\changed{For 21\% of the games we identified {\bf Neural Network as Teammate}, which happens if the interaction between the player and the NN is structured as those between colleagues.} In these games, the player and the NN work together toward a shared goal. For example, in \textit{Evolution} \citeL{evolution}, players and the NN create a stick-figure-like creature together. Players assemble the creature by placing bones, muscles, and joints in different ways. The NN takes the player's creation and improves it through evolving it over many iterations. This interaction creates a collaborative cycle between the player and the NN. A unique characteristic of this interaction metaphor is that the player and the NN have complementary skills. Both are needed to complete the game objective.
%


\changed{The final 19\% of the games used the {\bf Neural Network as Designer} metaphor.} In these games, the NN acts as a creator and the player as its client. The NN generates new content~\citeL{lessin2015darwin,aidungeon} or customizes content based on the preferences of the player \citeL{risi2015petalz,hastings2009evolving}, usually determined passively through players frequently interacting with a particular game element. For example, in \textit{Petalz} \citeL{risi2015petalz}, players arrange and nurture a balcony of flowers, which are generated by an NN. The NN generates each flower (shape and color) based on the player's selection of flowers to breed or cross-pollinate. The NN extends the game's playability by creating flowers that match the preferences of the player. Notice that compared to NN as Teammate, the player here generally has less well-defined goals to accomplish with the NN. 
%

\begin{figure*}[t]
  \includegraphics[width=\textwidth]{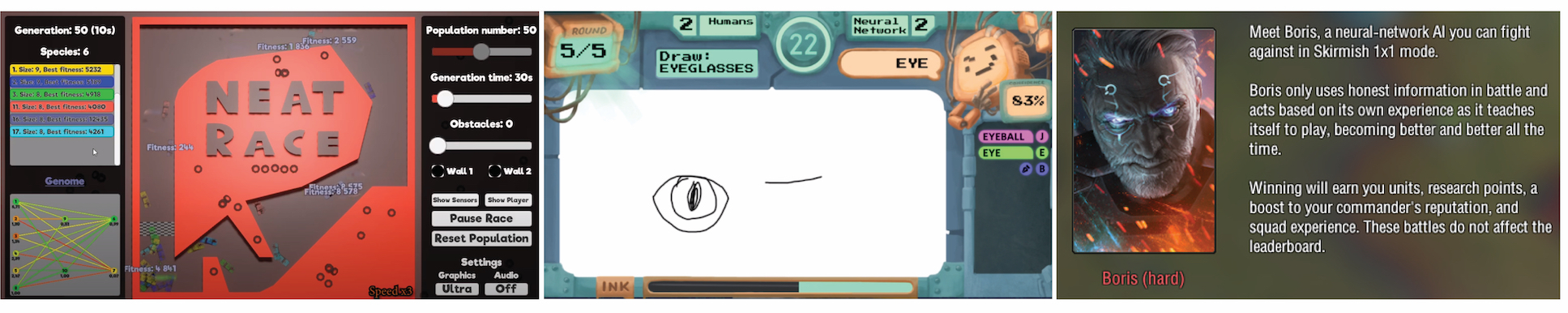}
  \caption{From left to right, we display \emph{Neat Race}~\protect\citeL{neat} categorized as \emph{NN-Specific}, {\em iNNk}~\protect\citeL{villareale2020innk} categorized as \emph{NN-Specific}, and {\em Blitzkrieg 3}~\protect\citeL{blitzkrieg} categorized as \emph{NN-Limited}.}
\label{fig:uivisibility_examples}
\end{figure*}

\subsubsection{Visibility of NN in Core UI}
\label{sec:visibility}

For the second research question of ``how visible are NNs in the UI of the core gameplay,'' we found 3 levels in which the NN is called into the player's attention in the UI: NN-Specific, NN-Limited, and NN-Agnostic.


\changed{A significant number of games (26\%) foregrounded the existence of its NN through what we call {\bf NN-Specific UI}.} These UIs highlight the presence of the NN during core gameplay through linguistic features (e.g., using the term ``neural network'' \citeL{snake,grid,idle,villareale2020innk}). For instance, \textit{How to Train your Snake} describes each NN-controlled snake as ``...hooked up to a Neural Network'' \citeL{snake}. Some games use visual features (e.g., visualizing the underling NN~\citeL{snake,idle,neat,football,villareale2020innk}). In {\em iNNk} (Middle, Figure~\ref{fig:uivisibility_examples}), the word ``neural network'' is prominently featured in the core game UI along with the NN's confidence meter. More interesting, some games visualize the parameters of the NN training algorithm to make the training process playable. For example, in {\em Neat Race}~\citeL{neat} (Left, Figure~\ref{fig:uivisibility_examples}), the game visualizes the NN's internal structure (bottom left of the screenshot) and displays its parameters as sliders (top right). 

\changed{The majority of our games (40\%) used {\bf NN-Limited UI}.} They acknowledge the presence of the NN in the game, but only through non-essential UI, such as using technical terminology in tutorials~\citeL{aidungeon,grand1997creatures,aivolution}, menus outside the core gameplay loop~\citeL{lessin2015darwin,forza,stanley2005evolving,grid}, or explicitly referring to the NN only in title screens~\citeL{semantris,aidungeon}. 
For instance, {\em Blitzkrieg 3} \citeL{blitzkrieg} is a WWII strategy game where players build and command a variety of units to defeat the opposing NN-controlled enemy. The game's opening screen (Right, Figure \ref{fig:uivisibility_examples}) personifies the NN as an evil-looking person with the text ``Meet Boris, a neural-network AI you can fight against...'' 

Finally, \changed{34\% of the games used {\bf NN-Agnostic UI}, which does not reference the NN.} 
By masking the NN, these games maintain the narrative immersion of the game worlds without revealing the algorithms used to build them.

\begin{figure}[!t]
 \includegraphics[width= \linewidth]{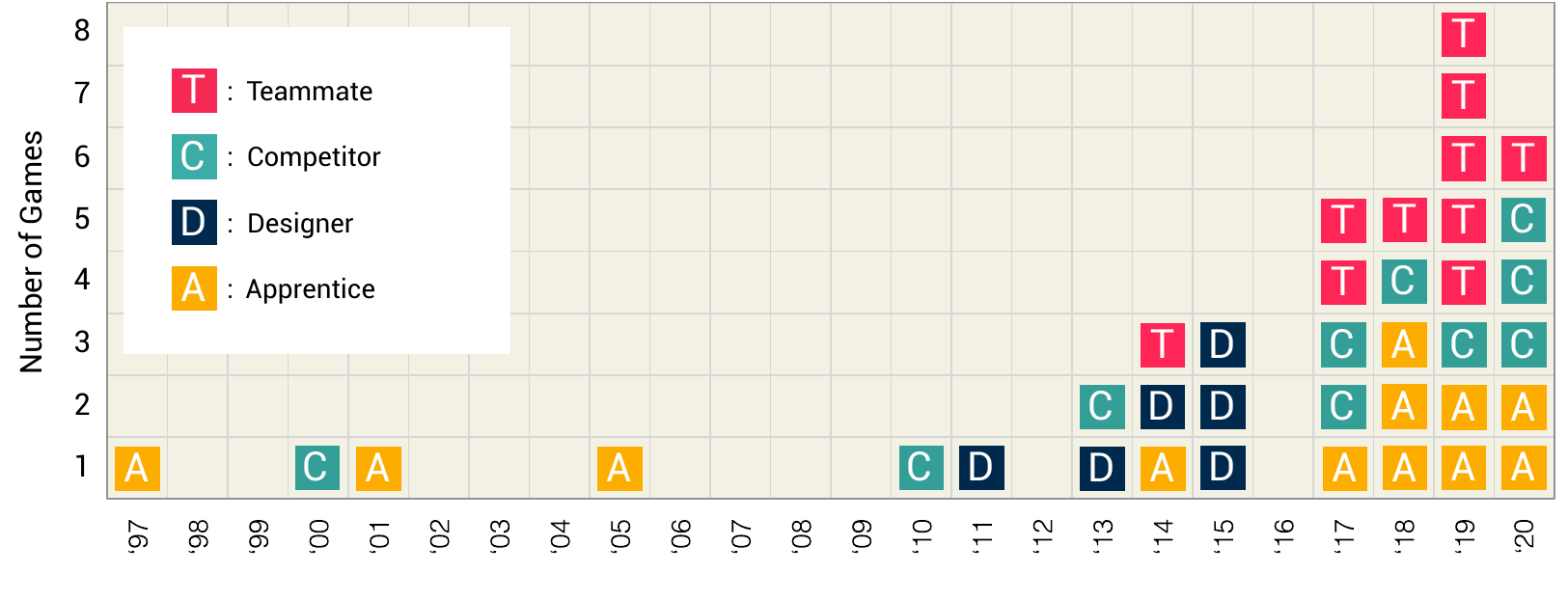}
 \caption{\changed{Distribution of the NN games by publishing date.}} 
\label{fig:games_year}
\end{figure}

\subsection{Discussion}



\subsubsection{NN Is Driving the Experimentation of Novel Gameplay Experiences} 
We observed a surge of NN games in recent years (Figure~\ref{fig:games_year}). NNs have been adopted in a wide variety of game genres and gameplay experiences. The games in our collection covered all the common game genres, showing vibrant efforts in trying different ways of incorporating NN in games. The interaction metaphor with the longest history is NN as Apprentice, whereas NN as Teammate has only been published since 2014. Our hypothesis is that the metaphor of NN as Teammate has the highest technical requirement for the NN because it needs to sufficiently understand and anticipate player interaction in order to accomplish a common goal. Thus, they are the most recent interaction metaphor. Most of the games are experimental, as the game design community has not formed established ways to use NNs. This is also echoed by the fact that indie developers and researchers developed the majority of our games (74\%). 

We noticed that some of the NN games are providing novel gameplay experiences that would not have been possible without NNs. For example, thanks to advancements in deep learning, {\em AI Dungeon}~\citeL{aidungeon} offered a text-based, human-AI collaborative storytelling experience with few constraints for what the player could type into the game. This is a drastic improvement over traditional text-based adventure games, which are infamous for their intolerance for player input that is slightly different from what the designer programmed for~\cite{montfort2005twisty}. More significant is that no matter what story elements the player enters, {\em AI Dungeon} can respond in reasonable ways. 

The NN games in our dataset have improved established gameplay experiences that had been supported by other AI techniques. Compared with traditional AI, NNs are more adaptive and flexible. In NN as competitor games, we observed that these games use established game mechanics (e.g., racing~\citeL{forza,moto}), but the NN offers a more challenging competitive experience through more capable NPCs than other AI techniques. 

The most unique gameplay is found in making the training of the NN itself a playable mechanic. By representing the NN and displaying features of its internal processes, this may lead to new player experiences, such as using gameplay to inform their understanding of the system to be successful in the game. For example, we observed that this occurs in games such as \textit{How to Train your Snake}~\citeL{snake} and \textit{Idle Machine Learning Game}~\citeL{idle}, which made the NN explicit in the UI and displayed parameters for the players to use when steering the NN's output. The gameplay became a puzzle about how to best configure the NN to be successful in the game. To do so requires players to have a basic understanding of the capabilities and limitations of the system, which are discovered over time through play.



\subsubsection{Metaphors We Play By}
In our analysis, we saw that interaction metaphors based on human relationships played a powerful role in structuring player-AI interaction. We did not notice that any games break or even complicate (e.g., a teammate that back-stabs) the interaction metaphor they use. Our empirical analysis validates what prior researchers proposed based on intuitions and domain knowledge~\cite{treanor2015ai,cook2016pcg,riedl2013ai,guzdial2019friend}, all of which uses human relationships. The role of metaphors has been extensively studied in human cognition~\cite{lakoff2008metaphors,lakoff2009more} and in UI design~\cite{Sharp2019,lubart2005can}, as it has been more recently in human-AI interaction as well~\cite{dove2020monsters}. 

We believe this full, uncomplicated adoption of interaction metaphors reflects the early stage of human-AI interaction. While we did notice some laudable innovations such as those mentioned in the previous section, by and large, game designers and developers went with familiar concepts and metaphors. This is consistent with Bolter's notion of how new technology {\em remediates} familiar forms before taking on its distinctive forms~\cite{bolter2016remediation}. Compared to ML-based UX~\cite{dove2017ux}, an advantage of the games community is that game AI developers are often game designers themselves or work closely with the latter. This cross-pollination between algorithm and design makes games a vibrant domain for new experimentation.

It is also notable that the metaphors are connected with the algorithmic characteristics of different NNs. As shown in Figure~\ref{fig:game_analysis_circlechart}, all games that adopted NN as Apprentice use online learning for player agency, whereas most games with NN as Competitor use offline learning for opponent competency.

\subsubsection{The Struggle with Transparency and Interpretability} 
Similar to most NN-based UX applications (see further discussions in Section 5), NN games struggle with how and what to communicate to the player regarding the use of the NN. \changed{In our initial coding stage, even our NN researcher coauthors could not always figure out the NN-specific questions by simply looking at the game. Because the use of NNs may not be apparent in the gameplay, and since only 26\% of our games references NNs (including simply using the word ``neural network'') in their core UI, the game requires the player to believe they are interacting with an NN.}  

Even when the players are directly playing with the parameters of the NN, it is not always clear what these features do. For example, unless the player has a background in evolutionary algorithms, terms such as ``population'' and ``retraining'' are not necessarily understandable.  

Like other AI-infused systems, games also struggle with the lack of interpretability of NNs. We see a full spectrum from completely blackbox NNs \citeL{grand1997creatures,blackandwhite,rabin2015game,democracy} (often in NN as competitors) to attempts to visualize the underlying NN \citeL{idle,football,snake,braincrafter,neat,stanley2005evolving}. Most notably, {\em NERO} \citeL{stanley2005evolving}, gives insights about the NN and its training in two ways. First, by visualizing the NN training parameters, players can steer the NN behavior by tweaking the reward structure for what is preferred behavior (e.g., approach enemy, attack enemy). Second, players are able to see graphs of the NN's internal structure and fitness values across generations for all robots across various combat stats (e.g., enemy hits).

The strongest designs for making NNs more interpretable come from simulation games where the player can tweak different training parameters. In these cases, even though the names of the parameters are sometimes too technical for players without an AI background, the NN's behavioral change feedback through different iterations of trial-and-error gameplay helps the player develop an intuition. In other words, most games in our dataset manage to reframe the difficulties of interacting with an NN as a puzzle and thus make it more engaging. 
\section{Phase 2: Analyzing NN Games with General Human-AI Interaction Guidelines}


The second broad research question is to explore what neural network (NN) games can tell us about designing human-AI interaction. Here we focus on the following subsidiary research questions:  

\begin{itemize}
    \item RQ 2.a: To what extent do NN games comply with contemporary design guidelines for human-AI interaction? 
    \item RQ 2.b: Using the design guidelines for human-AI interaction, how can NN games be differentiated according to their characteristics (see Table~\ref{tab:game_analysis_mainTable}) and in comparison with other AI-infused products? 
\end{itemize}


\subsection{Methods}
For inferring what NN games can tell us about human-AI interaction, we used the human-AI design guidelines proposed by Amershi et al.~\cite{amershi2019guidelines}. It is the most recent and comprehensive manner in which the design for human-AI interaction is documented thus far by the HCI community. As discussed above (Section~\ref{sect:RelatedWork_games}), currently no equivalent guidelines exist specifically for games. There are 18 guidelines in ~\cite{amershi2019guidelines} in total. They are grouped according to when the user is interacting with the AI: 1) initially, 2) during, 3) when wrong, and 4) overtime. The analysis procedure in adapting these guidelines to the NN games involved a three-step process: Step 1. defining guiding questions for NN games, Step 2 establishing codes for analyzing games, and Step 3 analyzing the games with Step 1 \& 2. Two researchers performed all steps in close coordination and checked the outcomes of each step with the other authors for verification and to reach consensus~\cite{hill2005consensual,richards2018practical}.   

\subsubsection{Step 1: Guiding Questions for NN games.}
Two researchers completed a detailed reading of each guideline to understand the guideline in the context of the original AI application examples (e.g., recommender systems, activity devices, etc.). Then, both researchers explored how the guidelines may be applied in the context of games. This process led to the definition of a question for each guideline to help orient the researchers when observing the games in Step 3. For example, for Guideline 15 ``encourage granular feedback,'' we defined the question, ``how do players indicate their feedback such as preference to the NN during gameplay?'' Table~\ref{tab:codes} shows all the original guidelines and our associated ``Guiding questions for NN games.''       


From this process, we agreed that guidelines in the ``when wrong'' category did not apply in the context of games. Games handle failure differently than other AI products, where failure is expected and, in fact, part of the main interaction and resulting experience~\cite{juul2013art,anderson2018failing}. 
Additionally, for AI-infused products, humans are consumers of the AI. By contrast, players in many of our games actively control how the AIs are trained. This close relationship significantly complicates the notion of failure in games. As a result, fully unpacking what failure means in games is out of the scope of this paper. We hence excluded this category and focused our analysis on the initially, during, and overtime categories. We do offer some observations in the discussion, but further research is needed in this important area of player-AI interaction.

\subsubsection{Step 2: Codes for Analyzing NN Games}
Two researchers took an iterative approach to arrive at a set of codes to analyze the games in Step 3. Amershi et al.'s~\cite{amershi2019guidelines} do not necessarily specify or recommend \textit{how} the guidelines should be evaluated, but in their user study with UX designers ({\em n} = 49), they applied a 5-point semantic differential scale from ``clearly violated'' to ``clearly applied.'' We combined their approach for heuristic evaluation with our guiding questions to create a 3-point coding scheme for each guideline. In a nutshell, this coding scheme is used to decide whether a game (A) clearly applies, (B) partially applies, or (C) violates the guideline. For example, for Guideline 1 ``Make clear what the system can do'' we added the guiding question ``How does the game make clear what the NN can do?'' and the following three codes: (A) Makes the NN's capabilities known; (B) Makes part of NN's capabilities known; and (C) Does not make the capabilities known at all. With this coding scheme, the researchers were able to clearly label in the context of a specific guideline and NN games. Additionally, the 3-point scale is much more suitable for qualitative/consensus coding. We aimed to \textit{describe} rather than only rate or score the games to extract meaningful insights, hence why we defined the 3-point coding scheme for each guideline in accordance with the associated guiding question. Table~\ref{tab:codes} shows the resulting codes. 

\subsubsection{Step 3: Analyzing NN Games}
Two researchers applied the codes from Step 2 for the analysis of the \changed{31 playable games from our corpus}, the results of which are presented in Section~\ref{sec:guide_results}. When analyzing each game, both researchers reviewed the data collected from Phase 1 (i.e., gameplay footage and written observations) and played the game. After reviewing this material, they independently assigned codes for each guideline per game. Disagreements were resolved through discussion. 

After the coding results were agreed upon, scores were assigned for cross-comparison with the characteristics found in Section~\ref{sec:phase1} (see Table~\ref{tab:game_analysis_mainTable}). We assigned a score of 2 for the clearly applies codes (A), 1 for the partially applies codes (B), and 0 for the violation codes (C). For the comparison of the guidelines (i.e., comparing G1 with G2, etc.), we then took the sum of scores of these resulting scores per guidelines and divided them by the maximum score per guideline to calculate the \% per guideline. For the cross-comparisons of the characteristics (on the interaction metaphors, visibility, developer, etc.), we normalized the sum of scores as we have a different number of games per characteristic and then took the average of the normalized sum of scores to calculate the \% per characteristic. 

In this section, we do not report the results on the characteristics of NN input and NN output as they do not provide any meaningful insights. We further omitted categories with a low number of cases (e.g., there is only one adventure game). Finally, for the comparison of our outcomes with other AI-infused products reported by Amershi et al.~\cite{amershi2019guidelines}, we considered the major patterns in the aggregate data for both (i.e., NN games vs. all other AI-infused products).

\begin{table*}[tb]
    \caption{Summary of the guideline analysis codes.}
  \includegraphics[width=0.8\textwidth]{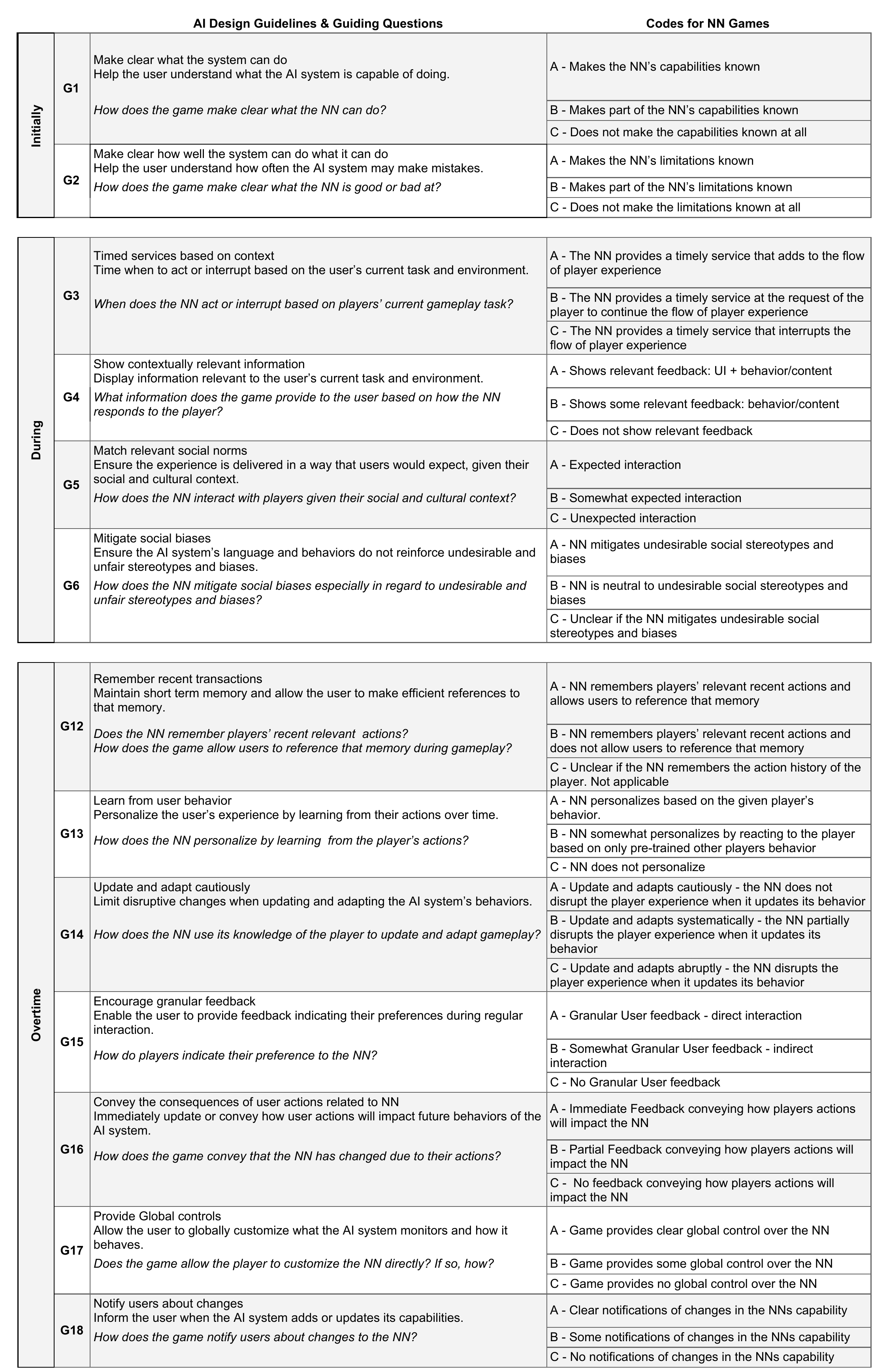}
\label{tab:codes}
\end{table*}

\subsection{Results} \label{sec:guide_results}

\begin{figure}
    \centering
    \includegraphics[width=0.8\linewidth]{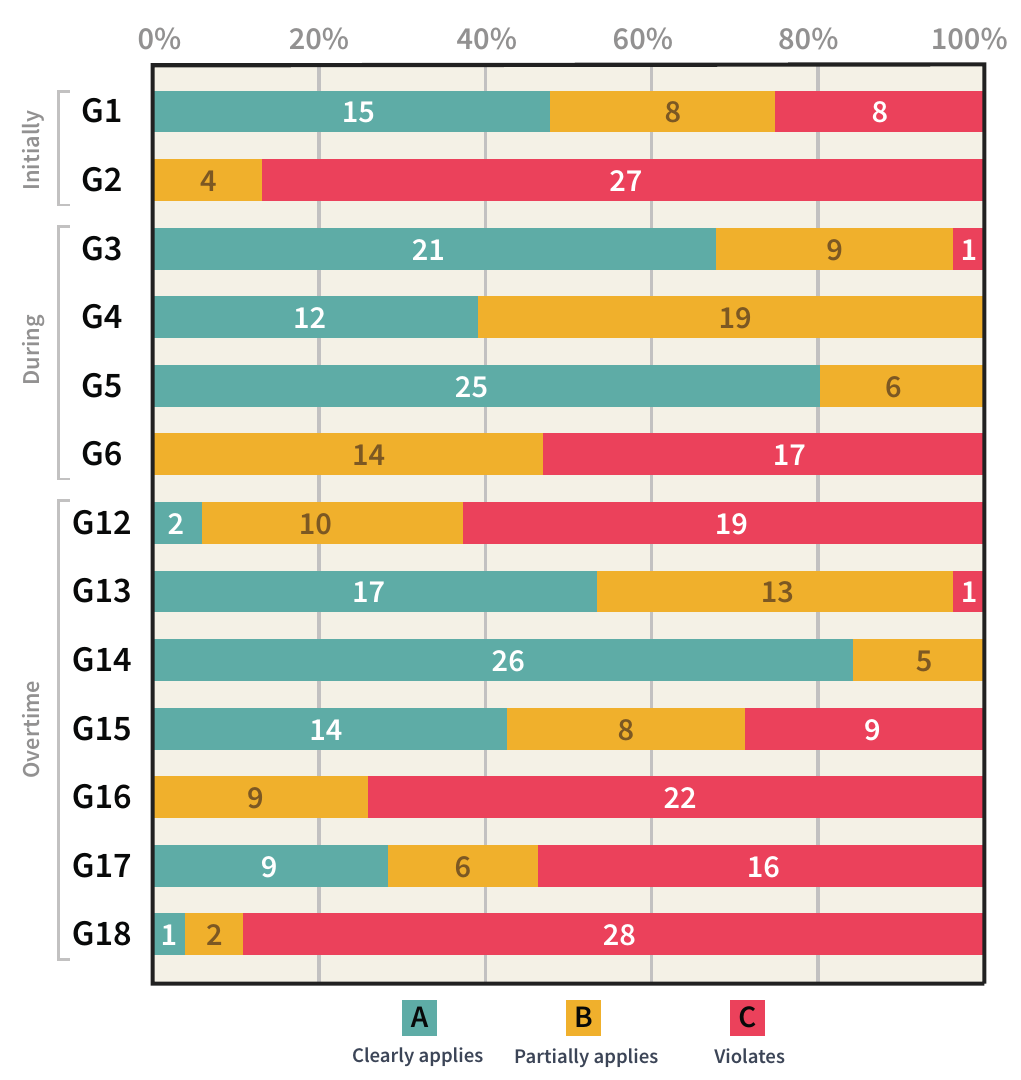}
    \caption{The distribution of compliance codes (A, B, and C) per guidelines in \changed{count ({\em n} = 31)} and \%.}
    \label{fig:guideline_results_stacked}
\end{figure}

Figure~\ref{fig:guideline_results_stacked} shows the results of applying the adapted guidelines in the context of NN games. Below, we discuss per guideline category (i.e., initially, during, and overtime) the results in more detail. Note that the reported \% here are based on the \% of games that were coded as A = clearly applied, B = partially applied, and C = violation. Following this, we compare the application of the guidelines across the characteristics discussed in Section~\ref{sec:phase1}, and with other AI-infused products. The reported results here are based on the \% derived from the normalized scores.    


\subsubsection{Initially}
The \textit{initially} guidelines (G1--G2) refer to player interaction prior to gameplay. During the initial interaction of these games, Guideline 1 ``makes clear what the system can do'' has the most reported applications: \changed{23 games (74\%)} made either full or part of the NN's capabilities known to the player prior to gameplay. These games helped the player understand the capabilities in a variety of ways, such as tutorials, intro screens, or developer notes prior to gameplay. Some were not as direct and did not provide such textual content, but still made the capabilities known by immediately providing the player with an output to observe. For example, in \emph{How to Train your Snake}~\citeL{snake}, players start the game to find the NN already training the snakes to move and find food, thus, showcasing an immediate result.

While these games are doing well in most cases by communicating the capabilities of the NN during the initial interaction, they are doing poorly regarding communicating the limitations of NNs to the players. Guideline 2 ``makes clear how well the system can do what it can do'' had \changed{27 games (85\%)} not making the limitations of the NNs known to the player at all --- the second-highest number of violations among the guidelines. This may be justified in cases playing against the NNs (i.e., NN as Competitor) to avoid exploiting the system to win the game. Or in cases where the NN is the focus of gameplay (i.e., NN as Apprentice), limitations become part of the puzzle and understanding through play.

\subsubsection{During}
The \textit{during} guidelines (G3--G6) refer to player interaction at any given gameplay loop. Guideline 3 ``time services based on context'' and 5 ``match relevant social norms'' had the most reported applications: \changed{21 games' (68\%)} NNs provided timely service, and \changed{25 games (81\%)} were labeled as an expected interaction with the NN. The clear application of G3 and G5 suggests how designers carefully considered how the NN can help assist with the flow of the player experience by suggesting interactions and providing immediate consequences that are consistent with player expectations.   

Games also performed well in complying with Guideline 4 ``show contextually relevant information,'' as none of the games violated showing contextually relevant information. In the majority of these games, the focus of the gameplay centers on changing or affecting the output of the NN. We observed that the games provide information in regards to how the NN was responding to or utilizing player actions. A common approach is the use of UI elements (e.g., NN performance stats increasing) accompanied by a continuous animation or visual change to enable players to observe and then inform their next gameplay action. For example, in \textit{iNNk} \citeL{villareale2020innk}, players can observe the exposed confidence meter that displays as a percentage under the NPC character in relation to their drawing, thus building a better mental model of how to subvert the NN in future drawings.


Other games provide additional visual information during this animation, such as an overlay of the entire NN population. Providing this extra information allows players to assess the NN's progress and observe both successful and failed attempts. For example, in \textit{Evolution} \citeL{evolution}, players are able to see an overlay of all the NN attempts at training the same creature simultaneously. This additional animation shows the highest and the lowest-performing creatures training at the same time, which enables players to better understand the progress as a whole and determine if the creature needs to be tweaked further.

While games are doing well to display relevant information regarding gameplay, Guideline 6 ``mitigate social biases'' had \changed{17 games (55\%)} with a violation. Examination of these instances revealed that it was unclear if the NN mitigates undesirable social stereotypes and biases. Additionally, we reported half of such games as a violation because such biases may emerge directly from players and are not mitigated. For example, games that made the NN the focus of gameplay (i.e., training or evolving using an NN) provides a new NN to play with. Therefore, players may steer the NN with their own personal preferences. Stereotypes and biases can emerge through the player's direction and reinforcement.

\subsubsection{Overtime}
The \textit{overtime} guidelines (G12--G18) refer to player interaction with the NN over a longer period of time. Guideline 13 ``learn from user behavior'' and 14 ``update and adapt cautiously'' are performing well. Based on the player's behavior, \changed{17 games (55\%)} NN personalize the experience, and \changed{26 games (84\%)} do not disrupt the gameplay experience when the NN changes its behavior. 
Guideline 15 ``encourage granular user feedback'' is doing well with \changed{14 games (45\%)} that allow players to directly indicate their preferences to the NN during gameplay. Games are performing moderately in regards to Guideline 17 ``provide global controls'' with \changed{15 games (48\%)} providing full or partial global control to adjust how the NN behaves, respectively.

A common approach to provide players more agency is the ability to adjust the NN through parameters or the environment it interacts in. We observed these in setting menus, or in other cases, directly in the core GUI. For example, in \emph{NERO}, the game allows players to edit the reward structure of the NNs during a training session. Further, players are able to edit the training environment, such as adding barriers and placing particular enemies to directly influence the NNs training.

Guideline G16 ``convey the consequences of user actions related to NN'' and G18 ``clear notifications of changes in NN capability'' had the most violations in this category: \changed{22 games (71\%)} did not provide any feedback conveying how players' actions will impact the NN, and \changed{28 games (90\%)} did not notify of any changes or updated to the NN capabilities. Further, Guideline G12 ``remember recent transactions'' was another difficult guideline to apply in games. \changed{In these cases, 10 games (32\%) leveraged the history of the player actions to generating content tailored to the player or a more challenging experience but did not allow the users to access that memory. Only 2 games (6\%) make this history useful to the player as other AI products do (e.g., navigation products, search engines) and allowed the users to reference that history.
}


\begin{figure}
\centering\begin{subfigure}[b]{\linewidth} 
\centering\includegraphics[width=7.5cm]{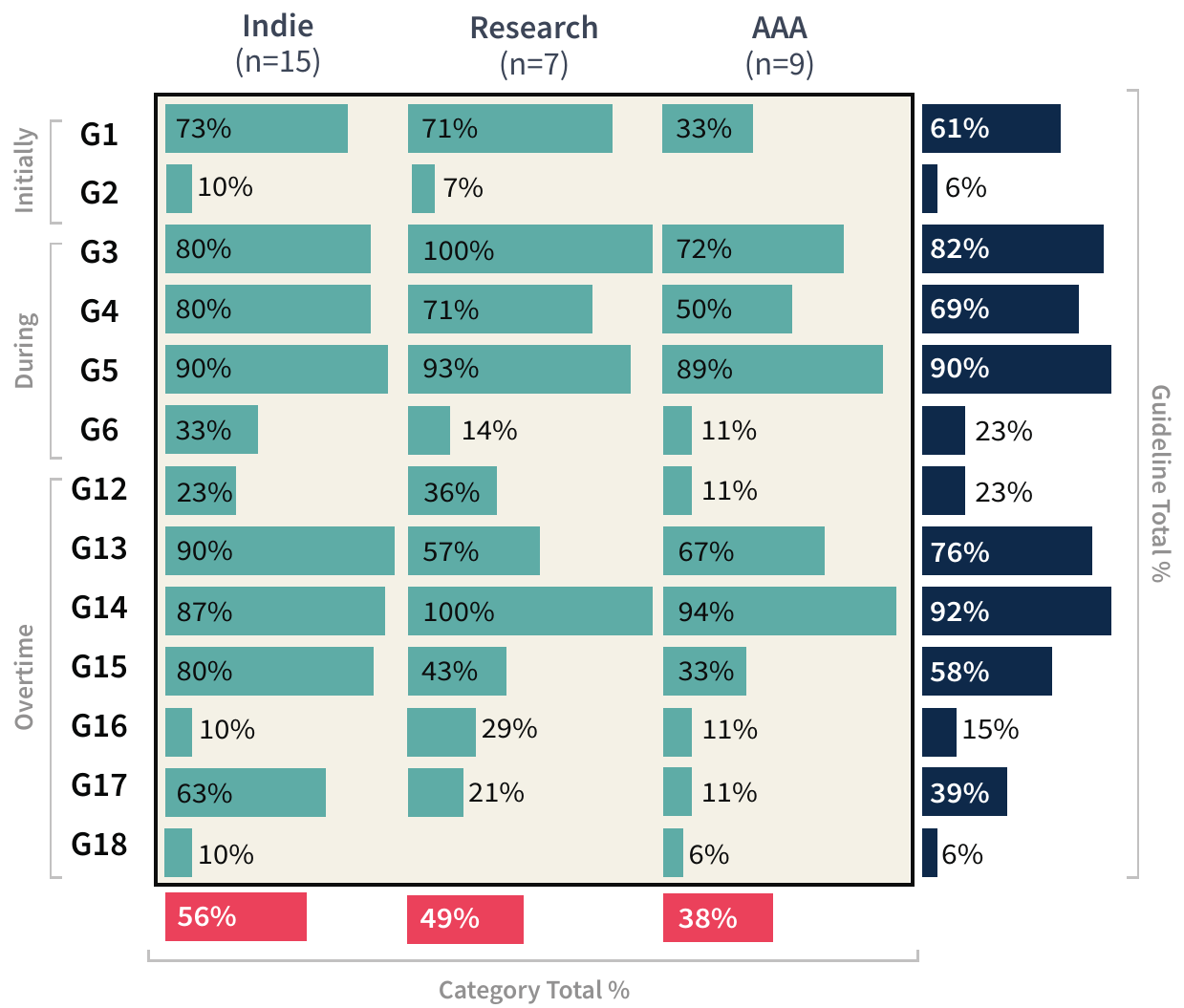} 
\caption{\label{fig:developer_score}Developer categories.} 
\end{subfigure}\vspace{10pt}

\begin{subfigure}[b]{\linewidth} 
\centering\includegraphics[width=7.5cm]{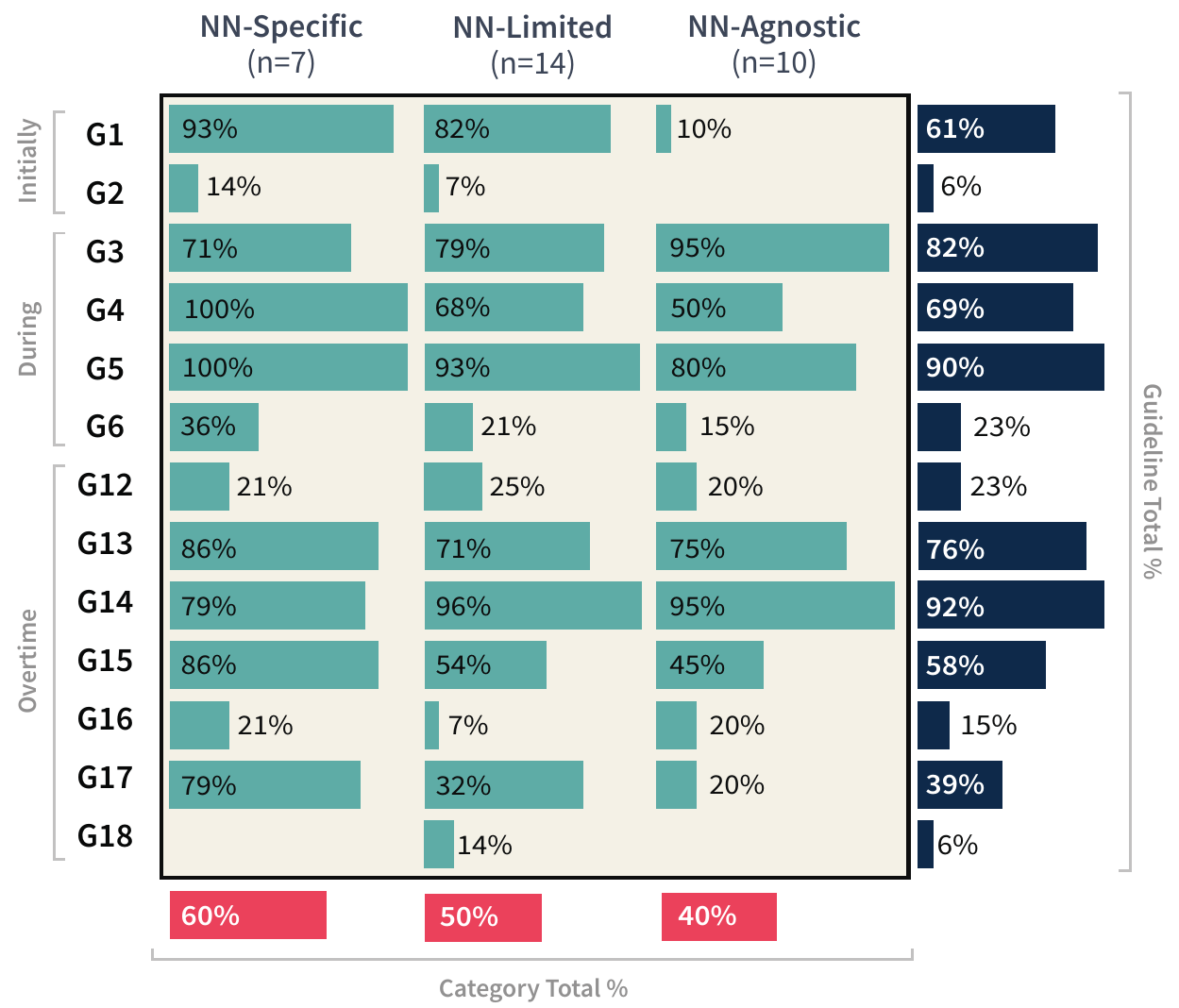} 
\caption{\label{fig:visibility_score}UI Visibility categories.} 
\end{subfigure}\vspace{10pt}

\begin{subfigure}[b]{\linewidth} 
\centering\includegraphics[width=8cm]{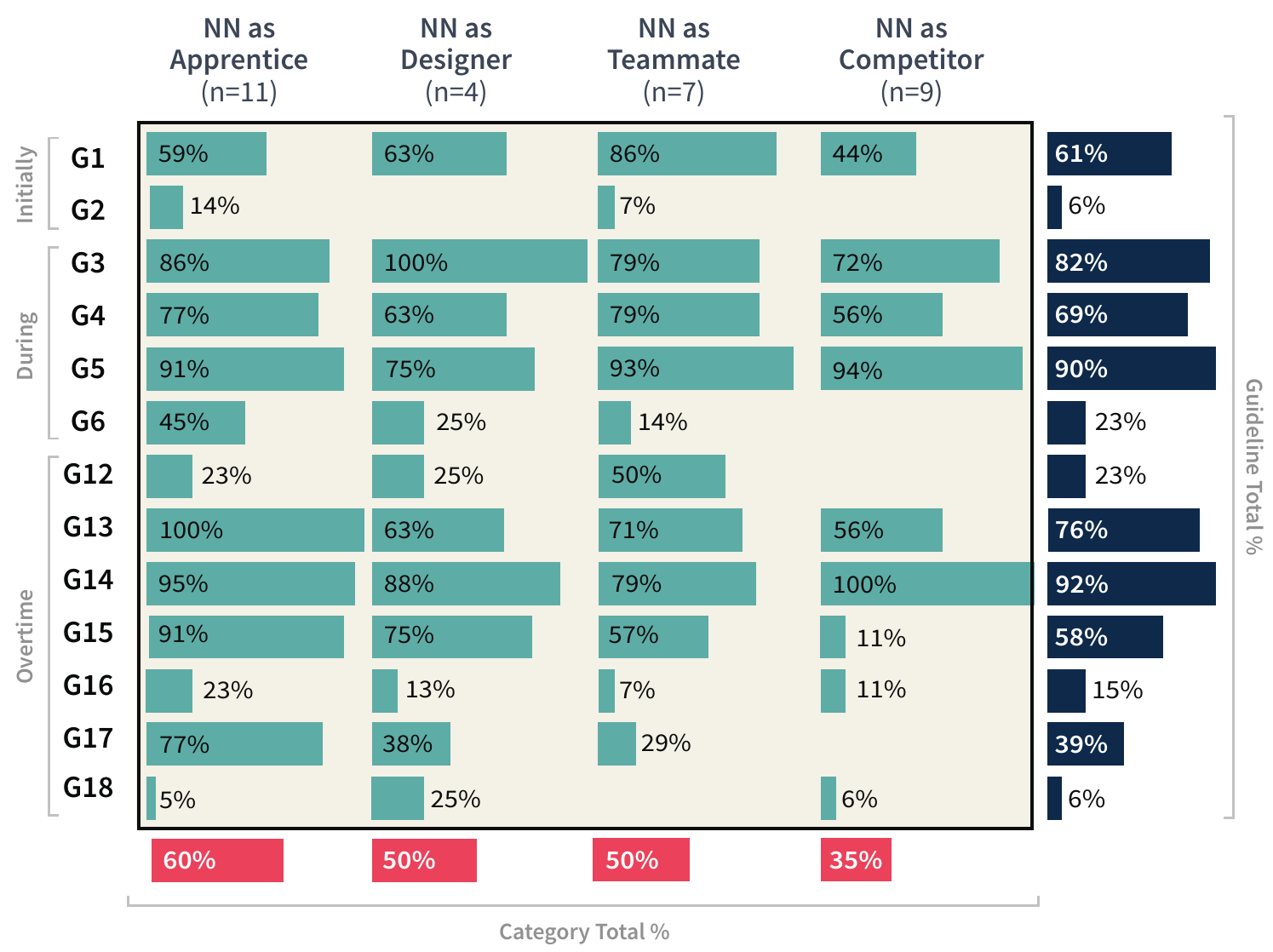} 
\caption{\label{fig:metaphor_score} Interaction metaphor categories.} 
\end{subfigure} 
\caption{A comparison of guidelines across different characteristics.} 
\end{figure}

\subsubsection{Comparison}
Consistent with the analysis in Section~\ref{sec:phase1}, NN-Specific games (60\%) outperform NN-Limited (50\%) and \changed{NN-Agnostic (40\%)} as shown in Figure~\ref{fig:visibility_score}. G1 ``make clear what the system can do'' is understandably what separates NN-Agnostic from the other two UI categories and G17 ``provide global controls'' is the most differentiating guideline for visibility. Consistent with the analysis in Section~\ref{sec:phase1}, we find that the \changed{simulation genre complies better (63\%) compared to all others (32--58\%)}, due to G4, G15, and G17. Also, not unexpectedly, NN games with online learning score higher \changed{(57\%) than those with offline learning (39\%)}, specifically in the guidelines in the overtime category. 

With the interaction metaphors (see Figure~\ref{fig:metaphor_score}), we see that NN as Apprentice scores the highest \changed{(60\%)} with here too G17 as the most differentiating guideline, followed by G13 ``learn from user behavior.'' Most apparent is that NN as Competitor \changed{(35\%)} scores the lowest, which is not unexpected given that there are gameplay reasons to not abide by the guidelines. Of note is that NN as Teammate do reasonably well on G12 ``remember recent transactions,'' which makes sense given that players need to build trust with the NN to work with them. 

We further find that the AAA games score lower \changed{(38\%) compared to the indie (56\%) and research (49\%) games} (see Figure~\ref{fig:developer_score}). This is mostly due that 6 out of 9 AAA games are classified as NN as Competitor games. It is interesting that research games score high on G3 and G14, while indie games score high on G13, G15, and G17. It indicates that the research games are more focused on integrating the NN into the game flow, while indie games are more experimenting in how players can interact with the NN. 

Aside from contrasting the outcomes with the game characteristics, we compared our guideline outcomes with the reported outcomes of AI-infused products by Amershi et al.~\cite{amershi2019guidelines} to see what key similarities and differences exist. We find that making the limitations known (G2) is poorly addressed in both NN games as other AI-infused products. AI-infused products do tend to perform better on G12, which is not surprising given that this a feature that is critical for many recommender types of products. NN games, however, generally perform better on G15 and G17, suggesting that NN games facilitate granular user feedback through direct interaction and provide more controls to their users, respectively. 

\subsection{Discussion} \label{Phase2_Disucssion}

By leveraging the design guidelines for human-AI interaction by Amershi et al.~\cite{amershi2019guidelines}, our aim was to explore what NN games tell us about designing human-AI interaction. In this process, we also learned more about NN games themselves and were able to verify and confirm the findings reported in Section~\ref{sec:phase1}. In fact, after applying the guidelines, we see more specifically what NN games do and how they differ from other AI-infused products. Here we describe the main takeaways from analyzing NN games with general human-AI interaction guidelines, especially in terms of how the notion of AI as play can be used to inform human-AI interaction.  



\subsubsection{Learning AI and Its Limitations Through Play}\label{5.3_a}
Clearly communicating the affordances and limitations of a system is a long-established design principle~\cite{norman2013design} and confirmed in human-AI interaction~\cite{amershi2019guidelines,de2017they,furqan2017learnability}. While the NN games do overall relatively well in communicating what the NN does (G1), they do not in communicating what it does not (G2). Other AI-infused products performed equally poorly on G2~\cite{amershi2019guidelines}. Through our close analysis of the games as well as developers' notes, we noticed that NN games \textit{purposely} ignore G2 because the point is to \textit{learn the limitations through play}.

The developers of \citeL{aidungeon,quick,semantris} all explicitly describe their games as experiences to explore or test the limits of the NN. For example, \textit{Semantris} asked the players to ``play around '' and ``see what references the AI understands best.'' In such games, the NN cannot directly communicate the limitations of the system upfront without jeopardizing the gameplay --- because discovering the system's limitations is the gameplay. By framing the discovery of NN limitations as play, many NN games are able to foster a sense of curiosity, discovery, and accomplishment in players. 

We also saw creative ways of making AI's limitations part of the rule of play. Most notably, we saw a number of projects with online-learning adopted by the idle game genre, which has a built-in play-wait-play cycle~\cite{alharthi2018playing}. This game feature makes the technical requirement of waiting for the NN to finish the training part of the expected experience and uses the idle game's reward system to incentivize players to return to the game after waiting. 

\subsubsection{Highlighting Failure as Part of Play}\label{5.3_b}
As argued above, failures in games are more complex than in other AI-infused systems. While we removed the ``when wrong'' category of guidelines in our analysis, we noticed many interesting uses of failure in the player-AI interactions in the NN games. Overall, failures in games are used productively. In many NN games, failure is used to motivate the player to continue improving their AI. This is consistent with the use of failure in game design~\cite{juul2013art}. 

Most notably, we noticed that failure is highlighted, instead of minimized, from the beginning. When the player first starts the game, the snake they control dies in \textit{How to Train your Snake}~\citeL{snake}, or robots run to the edges of the arena instead of approaching the enemy in \textit{NERO}~\citeL{stanley2005evolving}. This is a design pattern not commonly observed in non-AI games. We believe that the game designers used this device in order to re-frame players from ``problem makers'' to the AI into ``problem fixers'', making controlling the NN a less intimidating task. We noticed that this design strategy was used particularly in the NN as Apprentice games.

\subsubsection{Playing with Different forms of Human-AI Interaction}
Our analysis highlights the differences between various types of games, most notably in games using NN as Competitor. Compared to the other interaction metaphors, they violate many more guidelines. Through close examination, however, many of the violations are motivated and intentionally designed to be so. As argued above, when players compete with the AI, common design assumptions such as transparency and explainability do not apply directly and needs to be re-examined and adapted for the context. 

Literature on human-AI interaction primarily focuses on the paradigm of AI as a tool/augmentation to the user. However, an increasing number of AI-infused products fall outside this assumption. For example, AI in cybersecurity applications explores AI as an adversary, and AI products with high privacy concerns (e.g., in healthcare) require a different way to think about transparency. In these cases, we believe NN games can offer many design insights and cases for inspiration. We also need to expand current human-AI interaction guidelines so that it can encompass elements essential to play, such as engagement, flow, and fun.

\section{The Future of Player-AI Interaction}
Through the specific case of NN games, we have demonstrated the richness of player-AI interaction as a research topic. In this section, we intend to situate player-AI interaction in the broader context of related research areas. We then distill the design implications of our study for games and for UX/HCI. 

\subsection{Establishing Player-AI Interaction}
\begin{figure}
    \centering
    \includegraphics[width=0.6\linewidth]{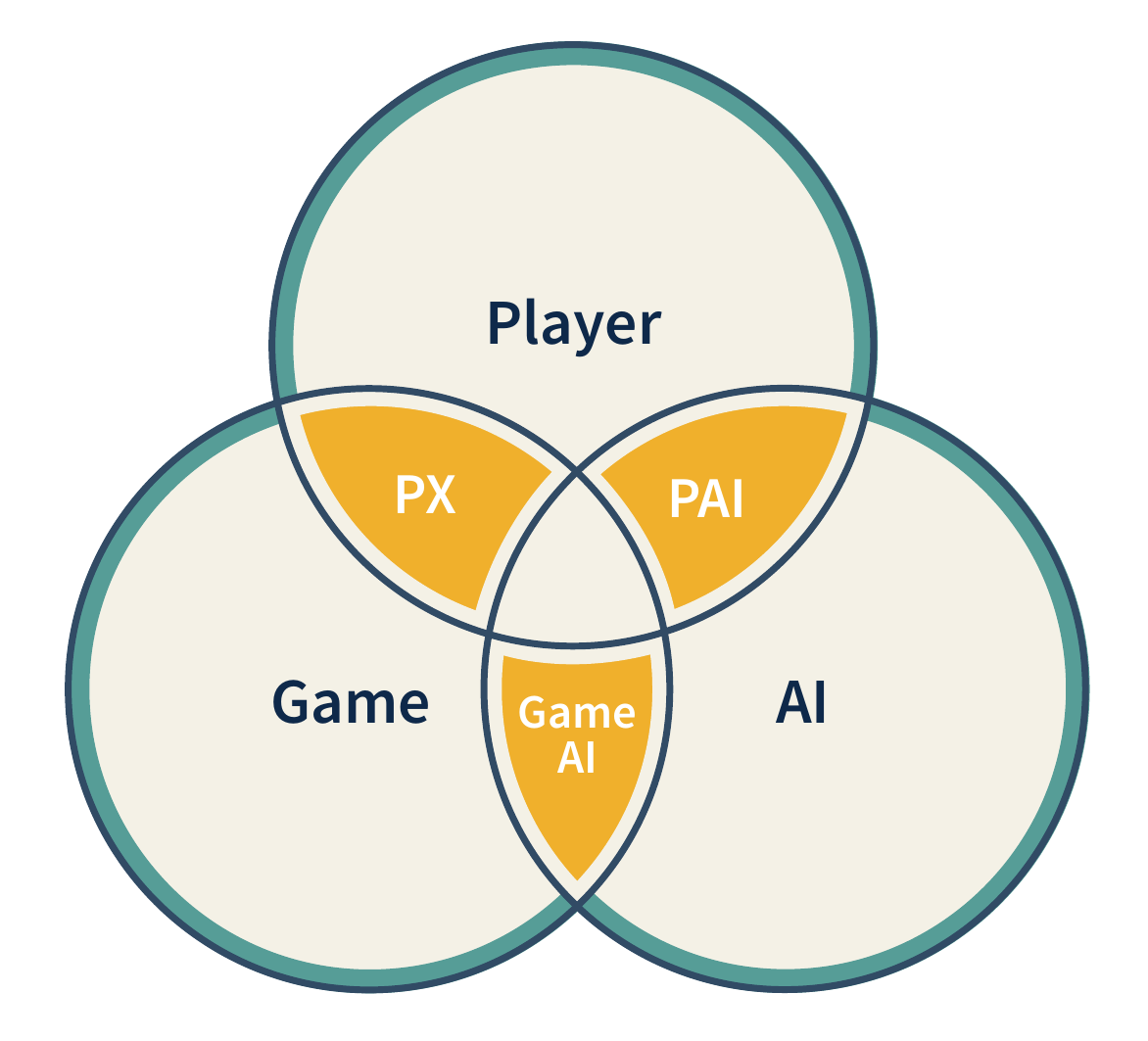}
    \caption{\changed{Research areas of games.}}
    \label{fig:researchAreas_vennDiagram}
\end{figure}

The study of players, game design, and game technology (here we focus on AI) are three main pillars in games research (Figure~\ref{fig:researchAreas_vennDiagram}). At the intersection of Games and AI, there is a well-established research community of game AI. In fact, game AI has often been driving the world of AI research, in which the most advanced forms of AI algorithms are often first developed and tested in games~\cite{risi2020chess}. Results in domains such as {\em StarCraft 2} and {\em Quake III} now frequently appear in the most prestigious journals~\cite{vinyals2019grandmaster,jaderberg2019human}. Combining the study of the player with game design, there is the research area of player experience. However, the intersection between players and AI has so far been relatively under-explored. Until recently, real players are typically only brought in at the end of game AI research as the means to evaluate the effectiveness of the algorithms. Carving out the topic of player-AI interaction will fill in this gap. 

In this paper, we present the first empirical study on this gap that we call \textit{player-AI interaction}. In addition to establishing games as a rich domain for human-AI interaction, our analyses contribute insights into how we can classify AI-based games based on interaction metaphors (i.e., apprentice, designer, teammate, and competitor) and the visibility of the AI system as part of the core UI (i.e., specific, limited, and agnostic). We further adapted design guidelines for human-AI interaction to the context of games, which can be useful for others when designing their AI-based games. However, we encourage the community to further scrutinize these design guidelines as our work indicates that AI-based games are different from AI-infused products, for example, with regards to the role of failure, and to work towards more formalized and evaluated design guidelines for player-human interaction.

\subsection{Encouraging Playing with AI}
For game designers, UX designers, and HCI researchers interested in human-AI interaction, one of our key takeaways is that reframing {\bf AI as play} offers a useful design approach, complementary to the current instrument-based views on AI. As as play can offer new human-AI interaction design space where the users can tinker, explore, and experiment with AI. We propose the following \textit{design considerations}:

{\bf Use flow to structure the learning curve of human-AI interaction}. For many users, interaction with AI can be overwhelming, especially when they encounter unexpected output from the algorithm. One important lesson from our study is that the concept of flow~\cite{csikszentmihalyi1990flow}, widely used to balance game difficulty and player engagement over time, can be useful to design human-AI interaction. In Section~\ref{Phase2_Disucssion}, we discussed that NN games should better support players for ``Learning AI and Its Limitations'' and make experimentation with AI more acceptable by ``Highlighting Failure as Part of Play.'' The use of flow can be useful to structure how to gradually expose users to different AI features (see also~\cite{cruz2017player}).

{\bf Incorporate enhanced discovery-based learning}. Many games in our analysis, especially simulation games, offer discovery-based learning~\cite{alfieri2011does} with mixed success. Since players come with different background knowledge and needs, explicit instruction for AI is challenging to design. Discovery-based learning offers players the opportunity to play around with the NN at their own pace and observe the consequences of their actions on the NN and the game world. However, most NN games in our dataset offered very little scaffold, making it difficult for players without a technical background to succeed. We suggest that UX designers use enhanced discovery-based learning and provide feedback, worked examples, scaffolding, and elicited explanations to further assist their users.

{\bf Extend the invitation to play.} Finally, for researchers and designers interested in exploring new forms of human-AI interaction, we believe offering users an invitation to play can unleash their imagination and empower them to explore new ways to interact with even the same technology. As we can see from {\em Hey Robot!}, the magic circle of play turns the smart speaker user from the seeker of information to the provider. The voice assistant's inability to understand user command/intent is transformed from failure to perform to the source of fun.

\section{Limitations}
We recognize that several limitations impact the scope of our work. First, our study considered only NN games with specific tags on popular game platforms and textbooks. 
\changed{Our goal was to find a representative sample of salient NN games, not a comprehensive list. However, we acknowledge that we may have missed some relevant games.}
%
Second, we omitted the failure-related human-AI interaction guidelines. While we offered some related observations, further research is needed to study failure in games, separate from failure outside the context of play, and how it relates to player-AI interaction. 
%
Third, we used 3-point codes in our qualitative analysis. While it is appropriate for our purpose of the analysis, future research can adopt a more fine-grained analysis that can better distinguish the ``violations'' and ``does not apply.'' \changed{Finally, we do not have necessary (sufficient) technical information about how NNs are used in many commercial games. The blackbox nature of game AI limited our ability to conduct in-depth analyses of specific features of games (see ``Multiple AI?'' in Table~\ref{tab:game_analysis_mainTable}).}






\section{Conclusion}
We introduced the term player-AI interaction to study how human players interact with AI in the context of games. While we intend to situate it in the broader context of human-AI interaction, we also highlight the unique opportunities and challenges presented by re-framing AI as play. Through a systematic search of existing neural network games, we conducted two deep qualitative analyses. In the first one, we analyzed the common metaphors that structure the player-AI interaction and how much the NNs are foregrounded in the core UI. In the second analysis, we adapted the current human-AI interaction guidelines to player-AI interaction and applied them to identify the strengths and weaknesses of NN games. Based on our findings, we proposed that the notion of AI as play, which is an alternative to the current paradigm of performance-centric human-AI interaction, can contribute to both game design and HCI communities.

\begin{acks}
This work is partially supported by the National Science Foundation (NSF) under Grant Number IIS-1816470 and a  DFF-Danish ERC-programme grant (9145-00003B). The authors would like to thank all past and current members of the project, especially Evan Freed and Anna Acosta for assistance in collecting initial data. We want to thank those who suggested additional games on {\em Twitter}. Finally, we thank Robert C. Gray for assistance in editing this paper. 
\end{acks}

\bibliographystyle{ACM-Reference-Format}
\bibliography{references}

\bibliographystyleL{ACM-Reference-Format}
\bibliographyL{references_games}
\nociteL{*}

\end{document}